\documentclass[twocolumn]{aastex63}

\def\arcsec{$^{\prime\prime}$} 
\newcommand{\msun}{$M_{\sun}$}
\newcommand{\rsun}{$R_{\sun}$}
\newcommand{\lsun}{$L_{\sun}$}
\newcommand{\msunyr}{\msun\,yr$^{-1}$}
\newcommand{\kms}{{\rm km\;s^{-1}}}
\newcommand{\escm}{{\rm erg\;s^{-1}\;cm^{-2}}}
\newcommand{\halpha}{H$\alpha$}
\newcommand{\hbeta}{H$\beta$}

\newcommand{\henir}{{\ion{He}{1}}$~\lambda$10830}
\newcommand{\mdot}{$\dot{M}$}
\newcommand{\ri}{$R_{\rm i}$}
\newcommand{\rw}{$W_{\rm r}$}
\newcommand{\tmax}{$T_{\rm max}$}

\def\curf{{\cal F}}
\newcommand{\vsini}{$v\sin i$}

\newcommand{\civ}{\ion{C}{4}}

\usepackage{graphicx}
\usepackage[caption=false]{subfig}
\usepackage{xcolor}
\usepackage{url}
\usepackage{booktabs}

\newcounter{column_number}
\shortauthors{Espaillat et al.}
\shorttitle{ODYSSEUS}

\begin{document}

\title{The ODYSSEUS Survey. Motivation and First Results: \\
Accretion, Ejection, and Disk Irradiation of CVSO~109}

\correspondingauthor{Catherine C. Espaillat}
\email{cce@bu.edu}

\collaboration{62}{}

\suppressAffiliations

\author[0000-0001-9227-5949]{C.~C. Espaillat}
\affil{Institute for Astrophysical Research, Department of Astronomy, Boston University, 725 Commonwealth Avenue, Boston, MA 02215, USA}

\author[0000-0002-7154-6065]{G.~J. Herczeg}
\affil{Kavli Institute for Astronomy and Astrophysics, Peking University, Yiheyuan 5, Haidian Qu, 100871 Beijing, China}
\affil{Department of Astronomy, Peking University, Yiheyuan 5, Haidian Qu, 100871 Beijing, China}

\author[0000-0003-4507-1710]{T. Thanathibodee}
\affil{Department of Astronomy, University of Michigan, 1085 South University Avenue, Ann Arbor, MI 48109, USA}

\author[0000-0001-9301-6252]{C. Pittman}
\affil{Institute for Astrophysical Research, Department of Astronomy, Boston University, 725 Commonwealth Avenue, Boston, MA 02215, USA}

\author{N. Calvet}
\affil{Department of Astronomy, University of Michigan, 1085 South University Avenue, Ann Arbor, MI 48109, USA}

\author{N. Arulanantham}
\affil{Space Telescope Science Institute, 3700 San Martin Drive, Baltimore, MD 21218, USA}

\author[0000-0002-1002-3674]{K. France}
\affiliation{Laboratory for Atmospheric and Space Physics, University of Colorado Boulder, Boulder, CO 80303, USA}

\author[0000-0001-7351-6540]{Javier Serna}
\affil{Instituto de Astronom\'{i}a, Universidad Aut\'{o}noma de M\'{e}xico,
Ensenada, B.C., Mexico}

\author[0000-0001-9797-5661]{J. Hern{\'a}ndez}
\affil{Instituto de Astronom\'{i}a, Universidad Aut\'{o}noma de M\'{e}xico,
Ensenada, B.C., Mexico}

\author[0000-0001-7157-6275]{\'A. K\'osp\'al}
\affiliation{Konkoly Observatory, Research Centre for Astronomy and Earth Sciences, E\"otv\"os Lor\'and Research Network, Konkoly-Thege Mikl\'os \'ut 15-17, 1121 Budapest, Hungary}
\affiliation{Max Planck Institute for Astronomy, K\"onigstuhl 17, 69117 Heidelberg, Germany}
\affiliation{ELTE E\"otv\"os Lor\'and University, Institute of Physics, P\'azm\'any P\'eter S\'et\'any 1/A, 1117 Budapest, Hungary}

\author[0000-0001-7796-1756]{F.~M. Walter}
\affil{Department of Physics and Astronomy, Stony Brook University, Stony Brook NY 11794-3800, USA}

\author[0000-0002-0474-0896]{A. Frasca}
\affiliation{INAF -- Osservatorio Astrofisico di Catania, via S. Sofia 78, 95123, Catania, Italy}

\author[0000-0002-3747-2496]{W.~J. Fischer}
\affil{Space Telescope Science Institute, 3700 San Martin Drive, Baltimore, MD 21218, USA}

\author[0000-0002-8828-6386]{C.~M. Johns--Krull}
\affiliation{Department of Physics and Astronomy, Rice University, Houston, TX  77005, USA}

\author[0000-0002-5094-2245]{P.~C. Schneider}
\affil{Hamburger Sternwarte, Gojenbergsweg 112, 21029 Hamburg, Germany}

\author{C. Robinson}
\affil{Department of Physics and Astronomy, Amherst College, C025 Science Center, 25 East Drive, Amherst, MA 01002, USA}

\author{Suzan Edwards}
\affil{Smith College, Northampton, MA 01063, USA}

\author[0000-0001-6015-646X]{P. \'Abrah\'am}
\affiliation{Konkoly Observatory, Research Centre for Astronomy and Earth Sciences, E\"otv\"os Lor\'and Research Network, Konkoly-Thege Mikl\'os \'ut 15-17, 1121 Budapest, Hungary}
\affiliation{ELTE E\"otv\"os Lor\'and University, Institute of Physics, P\'azm\'any P\'eter S\'et\'any 1/A, 1117 Budapest, Hungary}

\author[0000-0001-8060-1321]{Min Fang}
\affiliation{Purple Mountain Observatory, Chinese Academy of Sciences, 10 Yuanhua Road, Nanjing 210023, China}

\author[0000-0002-8476-1389]{J. Erkal}
\affil{European Southern Observatory, Karl-Schwarzschild-Strasse 2, 85748 Garching, Germany}

\author[0000-0003-3562-262X]{C.~F. Manara}
\affil{European Southern Observatory, Karl-Schwarzschild-Strasse 2, 85748 Garching, Germany}

\author[0000-0001-8657-095X]{J.~M. Alcal\'a}
\affil{INAF -- Osservatorio Astronomico di Capodimonte, via Moiariello 16, 80131 Naples, Italy}

\author[0000-0001-5260-7179]{E. Alecian}
\affil{Universit{\'e} Grenoble Alpes, CNRS, IPAG, 38000 Grenoble, France}

\author[0000-0001-6410-2899]{R.~D. Alexander}
\affiliation{School of Physics and Astronomy, University of Leicester, University Road, Leicester LE1 7RH, UK}

\author{J. Alonso-Santiago}
\affiliation{INAF -- Osservatorio Astrofisico di Catania, via S. Sofia 78, 95123, Catania, Italy}

\author[0000-0002-0666-3847]{Simone Antoniucci}
\affil{INAF -- Osservatorio Astronomico di Roma, via di Frascati 33, 00078, Monte Porzio Catone, Italy}

\author[0000-0002-2564-8116]{David R. Ardila}
\affiliation{Jet Propulsion Laboratory/California Institute of Technology, 4800 Oak Grove Drive, Pasadena, CA 91109, USA}

\author[0000-0003-4335-0900]{Andrea Banzatti}
\affiliation{Department of Physics, Texas State University, San Marcos, TX 78666, USA}

\author[0000-0002-7695-7605]{M. Benisty}
\affiliation{Universit{\'e} Grenoble Alpes, CNRS, IPAG, 38000 Grenoble, France}

\author[0000-0003-4179-6394]{Edwin A. Bergin}
\affil{Department of Astronomy, University of Michigan, 1085 South University Avenue, Ann Arbor, MI 48109, USA}

\author[0000-0002-1892-2180]{Katia Biazzo}
\affil{INAF -- Osservatorio Astronomico di Roma, via di Frascati 33, 00078, Monte Porzio Catone, Italy}

\author[0000-0001-7124-4094]{C\'esar Brice\~no}
\affil{Cerro Tololo Inter-American Observatory/NSF’s NOIRLab, Casilla 603, La Serena, Chile}

\author[0000-0002-3913-3746]{Justyn Campbell-White}
\affiliation{SUPA, School of Science and Engineering, University of Dundee, Nethergate, Dundee DD1 4HN, UK}

\author[0000-0003-2076-8001]{L. Ilsedore Cleeves}
\affiliation{Department of Astronomy, University of Virginia, Charlottesville, VA 22904, USA}

\author[0000-0002-2210-202X]{Deirdre Coffey}
\affiliation{School of Physics, University College Dublin, Dublin 4, Ireland}
\affil{The Dublin Institute for Advanced Studies, 31 Fitzwilliam Place, Dublin 2, Ireland}

\author[0000-0001-6496-0252]{Jochen Eisl\"offel}
\affiliation{Th\"uringer Landessternwarte, Sternwarte 5, D-07778 Tautenburg, Germany}

\author[0000-0003-4689-2684]{Stefano Facchini}
\affiliation{European Southern Observatory, Karl-Schwarzschild-Strasse 2, 85748 Garching, Germany}

\author{D. Fedele}
\affil{INAF -- Osservatorio Astrofisico di Torino, Via Osservatorio 20, I-10025, Pino Torinese, Italy}
\affil{INAF -- Osservatorio Astrofisico di Arcetri, Largo Enrico Fermi 5, 50125, Florence, Italy}

\author[0000-0002-5261-6216]{Eleonora Fiorellino}
\affiliation{Konkoly Observatory, Research Centre for Astronomy and Earth Sciences, E\"otv\"os Lor\'and Research Network, Konkoly-Thege Mikl\'os \'ut 15-17, 1121 Budapest, Hungary}
\affil{INAF -- Osservatorio Astronomico di Roma, via di Frascati 33, 00078, Monte Porzio Catone, Italy}

\author[0000-0003-4734-3345]{Dirk Froebrich}
\affil{School of Physical Sciences, University of Kent, Canterbury CT2 7NH, UK}

\author[0000-0002-8364-7795]{Manuele Gangi}
\affil{INAF -- Osservatorio Astronomico di Roma, via di Frascati 33, 00078, Monte Porzio Catone, Italy}

\author[0000-0002-7035-8513]{Teresa Giannini}
\affil{INAF -- Osservatorio Astronomico di Roma, via di Frascati 33, 00078, Monte Porzio Catone, Italy}

\author[0000-0001-5707-8448]{K. Grankin}
\affil{Department of Stellar Physics, Crimean Astrophysical Observatory, 298409 Nauchny, Crimea}

\author[0000-0003-4243-2840]{Hans Moritz G{\"u}nther}
\affiliation{MIT Kavli Institute for Astrophysics and Space Research, 77 Massachusetts Avenue, Cambridge, MA 02139, USA}

\author[0000-0003-0292-4832]{Zhen Guo}
\affil{Centre for Astrophysics Research, University of Hertfordshire, Hatfield AL10 9AB, UK}

\author[0000-0003-1430-8519]
{Lee Hartmann}
\affiliation{Department of Astronomy, University of Michigan, 1085 South University Avenue, Ann Arbor, MI 48109, USA}

\author{Lynne A. Hillenbrand}
\affiliation{Department of Astronomy, California Institute of Technology, 1216 East California Boulevard, Pasadena, CA 91125, USA}

\author[0000-0001-9504-0520]{P.~C. Hinton}
\affiliation{Laboratory for Atmospheric and Space Physics, University of Colorado Boulder, Boulder, CO 80303, USA}

\author{Joel H. Kastner}
\affiliation{Chester F. Carlson Center for Imaging Science, School of Physics and Astronomy, and Laboratory for Multiwavelength Astrophysics, Rochester Institute of Technology, Rochester, NY 14623, USA}

\author{Chris Koen}
\affiliation{Department of Statistics, University of the Western Cape, Private Bag X17,
Bellville, 7535, South Africa}

\author[0000-0001-8284-4343]{K. Mauc\'o}
\affil{Instituto  de  F\'isica  y  Astronom\'ia, Facultad  de  Ciencias,  Universidad de Valpara\'iso, Av. Gran Breta\~na 1111, 5030 Casilla, Valparaiso, Chile}
\affil{N\'ucleo Milenio de Formaci\'on Planetaria, Universidad de Valpara\'iso, Av. Gran Breta\~na 1111, Valparaiso, Chile}

\author[0000-0002-0233-5328]{I. Mendigut\'\i{}a}
\affil{Centro de Astrobiolog\'\i{}a (CSIC-INTA), ESA-ESAC Campus, 28692, Villanueva de la Cañada, Madrid, Spain}

\author[0000-0002-9190-0113]{B. Nisini}
\affil{INAF -- Osservatorio Astronomico di Roma, via di Frascati 33, 00078, Monte Porzio Catone, Italy}

\author[0000-0002-0151-2361]{Neelam Panwar}
\affil{Aryabhatta Research Institute of Observational Sciences, Manora Peak, Nainital, Uttarakhand 263001, India}

\author[0000-0002-7939-377X]{D. A. Principe}
\affiliation{MIT Kavli Institute for Astrophysics and Space Research, 77 Massachusetts Avenue, Cambridge, MA 02139, USA}

\author[0000-0002-9573-3199]{Massimo Robberto}
\affiliation{Johns Hopkins University, 3400 North Charles Street, Baltimore, MD 21218, USA}
\affiliation{Space Telescope Science Institute, 3700 San Martin Drive, Baltimore, MD 21218, USA}

\author[0000-0002-8421-0851]{A. Sicilia-Aguilar}
\affiliation{SUPA, School of Science and Engineering, University of Dundee, Nethergate, Dundee DD1 4HN, UK}

\author[0000-0003-3305-6281]{Jeff A. Valenti}
\affil{Space Telescope Science Institute, 3700 San Martin Drive, Baltimore, MD 21218, USA}

\author[0000-0002-6808-4066]{J. Wendeborn}
\affil{Institute for Astrophysical Research, Department of Astronomy, Boston University, 725 Commonwealth Avenue, Boston, MA 02215, USA}

\author[0000-0001-5058-695X]{Jonathan P. Williams}
\affil{Institute for Astronomy, University of Hawaii, 2680 Woodlawn Drive, Honolulu, HI 96822, USA}

\author{Ziyan Xu}
\affil{Kavli Institute for Astronomy and Astrophysics, Peking University, Yiheyuan 5, Haidian Qu, 100871 Beijing, China}
\affil{Department of Astronomy, Peking University, Yiheyuan 5, Haidian Qu, 100871 Beijing, China}

\author[0000-0002-6740-7425]{R.~K. Yadav}
\affil{National Astronomical Research Institute of Thailand, Sirindhorn AstroPark, 260 Moo 4, T. Donkaew, A. Maerim, Chiangmai 50180, Thailand}

\begin{abstract} 
	
The {\it Hubble UV Legacy Library of Young Stars as Essential Standards} (ULLYSES) Director's Discretionary Program of low-mass pre-main-sequence stars, coupled with forthcoming data from ALMA and {\it JWST}, will provide the foundation to revolutionize our understanding of the relationship between young stars and their protoplanetary disks. A comprehensive evaluation of the physics of disk evolution and planet formation requires understanding the intricate relationships between mass accretion, mass outflow, and disk structure. Here we describe the {\it Outflows and Disks around Young Stars: Synergies for the Exploration of ULLYSES Spectra} (ODYSSEUS) Survey and present initial results of the classical T Tauri Star CVSO~109 in Orion OB1b as a demonstration of the science that will result from the survey. ODYSSEUS will analyze the ULLYSES spectral database, ensuring a uniform and systematic approach in order to (1) measure how the accretion flow depends on the accretion rate and magnetic structures, (2) determine where winds and jets are launched and how mass-loss rates compare with accretion, and (3) establish the influence of FUV radiation on the chemistry of the warm inner regions of planet-forming disks. ODYSSEUS will also acquire and provide contemporaneous observations at X-ray, optical, NIR, and millimeter wavelengths to enhance the impact of the ULLYSES data. Our goal is to provide a consistent framework to accurately measure the level and evolution of mass accretion in protoplanetary disks, the properties and magnitudes of inner-disk mass loss, and the influence of UV radiation fields that determine ionization levels and drive disk chemistry.   
\end{abstract}


\section{Introduction} \label{intro}

The {\it Hubble UV Legacy Library of Young Stars as Essential Standards} (ULLYSES)\footnote{https://ullyses.stsci.edu} Director's Discretionary Program offers a once-in-a-lifetime opportunity to deepen our understanding of the connection between accreting young stars and their planet-forming disks. After emerging from opaque shrouds of dusty infalling gas, low-mass pre-main-sequence stars become optically visible as classical T Tauri stars (CTTS), revealing evidence for magnetospheric accretion from their disks. In the magnetospheric accretion paradigm \citep[see review by][]{hartmann16},  strong stellar magnetic fields truncate the inner disk at a few stellar radii \citep{donati09,johns-krull13}. Gas flows from the truncation radius along stellar magnetic field lines,  crashing onto the star and forming an accretion shock.  Accretion-powered winds and jets carry away angular momentum of the accreting gas, controlling the rotational evolution of both the star and the disk. The accretion shock produces strong ultraviolet (UV) and X-ray emission that irradiates the disk, affecting the chemistry of the planet-forming environment and enabling thermal winds that can regulate the disk lifetime. While images from the Atacama Large Millimeter/submillimeter Array (ALMA) of protoplanetary disks reveal dramatic structures thought to result from ongoing planet formation and early stages of disk dissipation \citep[see review by][]{andrews20}, resolving the innermost regions of these disks still remains beyond current capabilities.

Far- and near-ultraviolet (FUV and NUV) spectroscopy provides unique and powerful spectral diagnostics of young stars and innermost disk regions, conveying crucial information to help answer some of the fundamental questions about CTTS accretion flows, mass and angular momentum transport in the disk, and disk irradiation \citep[see review by][]{schneider20}. A number of questions arise. For accretion flows, how do the properties of the shock depend on the accretion rate and magnetic structures, which in turn depend on the mass of the star? For winds, how and where are they launched, and how do their mass-loss rates compare with mass lost in large-scale collimated jets and to accretion? For disks, what is the structure of the innermost regions, and how does the structure filter the chemistry-driving UV radiation incident on the disk, which alters the composition of forming planetary systems? Characterizing the flow of material and the accretion-generated UV radiation is essential to understanding the physics that controls the evolution of planet-forming disks.

Here we describe {\it Outflows and Disks around Young Stars: Synergies for the Exploration of ULLYSES Spectra} (ODYSSEUS), a comprehensive community effort to use the ULLYSES UV survey of CTTS to advance our understanding of accretion and outflow physics and disk evolution. The ULLYSES survey consists of $\sim500$ Hubble Space Telescope ({\it HST}) orbits dedicated to Cosmic Origins Spectrograph (COS) FUV spectra ($R\sim12,000$--$20,000$) and low-resolution ($R\sim500$) Space Telescope Imaging Spectrometer (STIS) NUV and optical spectra of $\sim60$--70 CTTS, covering a combination of stellar masses, accretion rates, and ages previously unprobed in the FUV and NUV. ODYSSEUS will provide a uniform, systematic analysis of the ULLYSES data, including the application of new, rigorous models for accretion shocks and winds, as well as extraction and interpretation of the complex forest of FUV lines. These analyses are interconnected, with interpretations that rely upon one another and upon a consistent set of star and disk properties. We are also acquiring, analyzing, and providing to the community extensive datasets of simultaneous and contemporaneous observations from complementary facilities (see Section~2) in the X-ray to millimeter wavelengths to enhance the impact of the ULLYSES data. These analyses will then be synthesized for use by the broader community, including as necessary inputs for interpreting James Webb Space Telescope ({\it JWST}) and ALMA observations.

We begin by introducing the background for FUV observations of accretion (\S 1.1), outflows (\S 1.2), and irradiated disks (\S 1.3), to place our results in context.

\subsection{Mass Accretion Via the Stellar Magnetic Field}

The magnetospheric accretion paradigm for CTTS is founded on the interpretation of excess UV and optical continuum emission as arising in the accretion shock \citep{calvet98,robinson19}, and on broad line profiles of hydrogen and low-ionization metals as forming in extended infalling magnetospheric flows \citep{muzerolle01,kurosawa06,alencar12}. The structure of the shock \citep[see Figure 2 of][]{hartmann16} attributes the excess continuum emission to a combination of regions, which forms the basis for extracting the mass accretion rate onto the star by fitting models to the observed continuum excess. 

The mass accretion rate is a fundamental physical property that has traditionally been derived by assuming uniform accretion columns with high-energy fluxes that dominate the UV and optical emission \citep{gullbring00}, but increasing evidence posits that multicomponent magnetic columns and energy fluxes are required to explain the excess emission over wide wavelength ranges \citep{donati11,ingleby13,schneider18,robinson19}. Physically, the presence of multicolumn accretion flows and shocks implies magnetospheric flows that may not follow a simple uniform dipole. Combining new multicomponent accretion shock models with spectra from the FUV to the near-infrared (NIR) will allow us to robustly measure mass accretion rates. UV spectra are particularly crucial to reveal accretion signatures in stars that are accreting at slow mass accretion rates \citep{ingleby11b,alcala19}. Accurately measuring accretion rates for CTTS in a statistically significant sample over a broad range of ages and stellar masses helps to constrain the timescale for stellar mass assembly and the onset of planet formation. 

The ``hot" FUV lines, \ion{Si}{4}, \ion{C}{4}, \ion{N}{5}, and \ion{He}{2}, complement the continuum diagnostics by probing hotter gas in the postshock and immediate preshock regions. The line luminosities correlate with accretion luminosity \citep[e.g.][]{ingleby11a,gomezdecastro12, yang12,robinson19} and have line profiles with a two-component structure: a narrow component produced in the shock and a broad component in the extended magnetospheric flows \citep{ardila13, gomezdecastro12}. However, no detailed simulations of these lines have been included self-consistently in existing accretion shock models. ODYSSEUS will analyze these hot lines with new state-of-the-art models that also explain the continuum excess. Accurate modeling of these hot lines will provide information about abundances in the accretion flow for evidence of depletion of silicates \citep{herczeg02,france10,booth18}, as seen in X-ray and some optical spectra \citep[e.g.][]{kastner02,kama15}, as a possible signature of advanced planet formation in the disk.

\subsection{Mass Outflow Via Winds and Jets}

Magnetohydrodynamic (MHD) winds may carry away most of the angular momentum of the disk, thereby allowing the gas to accrete and locking the stellar and inner disk rotation. Magnetic geometry determines how accretion-powered mass loss may be generated: primarily by magnetic phenomena tied to the star, by the interaction of the stellar magnetosphere with the inner disk, or by MHD winds from the inner disk itself. Each of these mechanisms plays a role at different radii; the shapes and strengths of observed wind tracers depend on the mass-loss rate and viewing geometry.

The most definitive signatures of winds come from blueshifted forbidden line emission and from blueshifted P~Cygni absorption in strong permitted lines. For example, both the forbidden line emission and P~Cygni absorption in \ion{He}{1} $\lambda10830$ trace a fast jet launched near the inner disk and a slower wind that is likely to arise from the disk itself over radii from 0.1 to 10~au \citep[e.g.,][]{edwards06,banzatti19}. The relationship between the winds probed by these P~Cygni absorptions and the fast collimated jet is not known. Differences in the physical interpretation of these winds \citep{dupree05,johns-krull07} have led to uncertainty in the mass fluxes and launch locations and therefore in their potential role in angular momentum transport.

The ionization balance in the wind is critical for determining whether the launch location is the inner disk, the magnetosphere--disk interaction region, or the star itself. It is also vital for measuring how much angular momentum is carried away. P~Cygni absorption in the FUV, in lines such as \ion{N}{1}, \ion{Fe}{2}, \ion{C}{2}, \ion{Si}{2}, \ion{Si}{3}, and \ion{C}{4} \citep{herczeg05,cauley16}, traces the same flows as does He I but probes a broad range of excitation and ionization (unique to the UV). This broad array of lines will break degeneracies in interpreting the \ion{He}{1} line alone. ODYSSEUS is developing MHD wind models tailored specifically to the FUV absorption lines to evaluate ionization and mass loss from column densities in absorption lines.

\subsection{The Structure and Chemistry of the Inner Planet-forming Disk Regions}

Over the past decade, surveys of CO, H$_2$O, and organic molecules have provided new constraints on the radial density, temperature, and composition profiles of the inner regions of planet-forming disks \citep[e.g.,][]{salyk11,carr11,banzatti17}. Extensive modeling of these species is being developed for ALMA and in anticipation of {\it JWST} \citep[e.g.,][]{semenov11,haworth16,bosman17}.  While these disk tracers are each well studied individually, a goal of ODYSSEUS is to provide a unified framework that combines and explains tracers of both the inner and outer disks, from the upper surface layers down to the midplane.  

In some cases, our line of sight to the star passes through the upper layers of the disk, leading to direct and powerful measurements of the CO/H$_{2}$ abundance ratio \citep{france14,cauley21}.  Disks with these geometries offer our best opportunity to solve one of the most perplexing, yet important, problems related to disks that have been identified with ALMA: the potential depletion of CO relative to H$_2$ \citep{mcclure16,miotello17}, with consequences for abundances of planets \citep[e.g.][]{oberg11,krijt18,aschneider21}. Since lines of sight that pass through the warm surface layers of the disk are rare (only $\sim10$ of $\sim50$ CTTS disks in the {\it HST}--COS archival data show strong CO absorption), ULLYSES will provide a larger sample of disks with CO absorption. Thanks to the high S/N requirements of the ULLYSES survey, we will measure CO column density and temperature of the inner disk gas with unprecedented precision; these observations create for the first time the ideal target set for dedicated follow-up to measure CO/H$_2$ ratios in a statistical sample of disks.

FUV molecular emission from disks complements mid-infrared (MIR) and submillimeter diagnostics: Irradiation by Ly$\alpha$ excites warm \citep[$>1500$~K,][]{nomura05,adamkovics16} H$_2$ and cool (200--500~K) CO gas, leading to strong emission lines in the UV. ODYSSEUS's analyses of warm H$_2$ and CO emission and absorption in the FUV will characterize the role of the UV radiation field, allowing us to connect the impact of this radiation on the disk. The survey will assemble high-resolution FUV spectra from ULLYSES, MIR spectra from {\it Spitzer} (and {\it JWST}, when available) for gas emission, and ALMA measurements of molecular lines to provide an indispensable large, cohesive and systematic dataset of planet-forming disks from stellar to solar-system scales.

In Section~\ref{data}, we discuss the data we expect to obtain to supplement the {\it HST} UV observations; this includes optical and NIR spectra, optical photometry, X-ray data, and potential future data with ALMA and {\it JWST}. We then present a first look at the results for the CTTS CVSO~109 in order to demonstrate the analysis we will undertake with ODYSSEUS for the entire ULLYSES sample. We begin with an overview of our observations of CVSO~109 in Section~\ref{obs}. In Section~\ref{sec:stellarproperties}, we derive the stellar and disk properties of the CVSO~109 system. In Section~\ref{sec:variability}, we look at optical light curves and spectra taken contemporaneously with the {\it HST} data and discuss their variability in order to place the {\it HST} observations into context to examine the {\it HST} and all other data together. We derive and discuss the accretion, ejection, and disk irradiation properties of CVSO~109 in Section~\ref{sec:accretion}. We end in Section~\ref{summary} with our summary and conclusions.

\section{Observations to Supplement ULLYSES} \label{data}

Since CTTS are highly variable, a robust set of complementary observations is needed to interpret the FUV and NUV probes of accretion, winds, and disks. ODYSSEUS is organizing contemporaneous and simultaneous observations with a wide range of facilities. Reduced high-level data products will be made publicly available, when possible with no proprietary period, eventually through the Mikulski Archive for Space Telescopes (MAST) and VizieR and in the short-term through Zenodo.\footnote{https://zenodo.org/communities/odysseus} 

\subsection{UV Data}

ULLYSES will spend about 500 {\it HST} orbits in cycles 28 and 29 (2020--2022) to obtain UV data of CTTS \citep{roman-duval20}. The survey sample consists of $\sim60$--70 CTTS with spectral types K through M in multiple star-forming regions with distances ranging from about 100 to 450 pc that will be observed once.  This component of the program samples CTTS with a broad range of masses and mass accretion rates, extending coverage in particular to lower masses than have previously been surveyed. The emphasis on southern star-forming regions is intended to complement a previous emphasis on Taurus in archival {\it HST} data. Four objects (TW~Hya, BP~Tau, GM~Aur, and RU~Lup) make up the monitoring sample and will be observed 24 times each. 
These objects will be observed four times per rotation period for three rotation periods; this pattern will be repeated again about a year later.


The survey targets will be observed with COS/G130M/1291 and COS/G160M/multiple cenwaves to provide continuous spectral coverage and with STIS/G230L, STIS/G430L, and STIS/G750L. The four monitoring targets will be observed with COS/G160M and COS/G230L. More details on the ULLYSES strategy and the CTTS sample can be found at the ULLYSES webpage.\footnote{https://ullyses.stsci.edu}

\subsection{Optical and NIR Spectra}

High-resolution optical spectra will be obtained of the survey and monitoring samples to accurately measure the stellar parameters and extinction at the time of the observations \citep{manara13b,alcala17,frasca17}. In addition, analyses of UV line profiles will be strengthened with contemporaneous constraints on funnel flows in high-resolution optical and NIR spectra (H series, \ion{He}{1} $\lambda10830$ line). The bulk of this data will be obtained from the ESO Very Large Telescope (VLT) through the Large Program PENELLOPE \citep{manara21}, which will observe the targets using the VLT/X-shooter, VLT/UVES (Ultraviolet and Visual Echelle Spectrograph), and VLT/ESPRESSO (Echelle Spectrograph for Rocky Exoplanets and Stable Spectroscopic Observations) spectrographs. We are also planning to obtain data with the National Optical-Infrared Astronomy Research Laboratory/CHIRON, Telescopio Nazionale Galileo/GIARPS, and, in select cases, with spectropolarimetry for magnetic field strength and topology, using an echelle spectropolarimetric device for the observation of stars (ESPaDOnS) and the infrared spectropolarimeter SPIRou at the Canada--France--Hawaii Telescope.

\subsection{Optical Photometry}  
While each {\it HST} visit includes an optical spectrum, accretion and extinction may change during $\sim6$ hrs of FUV spectroscopy. Photometric monitoring throughout each visit is needed to interpret the UV emission and to analyze variability within the time-tagged spectra. We are organizing global photometry efforts to obtain this data for the survey and monitoring samples, including the Las Cumbres Observatory Global Telescope (through the National Optical Astronomy Observatory), Konkoly Observatory (Hungary), the Osservatorio Astrofisico di Catania (Italy), and amateurs (through the American Association of Variable Star Observers [AAVSO, AAVSOnet]). The angular resolution of the optical images from Konkoly Observatory is limited by the seeing. From the measured FWHM of the Gaussians fitted to the science targets, we estimate that the seeing varied between 1.7 and 4.9$''$ during the observations, with 2--3$''$ being the most typical values. Many of the {\it HST} observations are also being scheduled to overlap with the Transiting Exoplanet Survey Satellite ({\it TESS}), which will provide broadband optical light curves with a cadence of at least 30 min.

\begin{deluxetable*}{cccccccc}
\tablecaption{Hubble Space Telescope Observations \label{tab:hst}}
\tablehead{
\colhead{Program} &
\colhead{Instrument} &
\colhead{Grating} &
\colhead{$\lambda_{\rm cen}$ (\AA)} &
\colhead{Wavelength Range (\AA)} &
\colhead{MJD at Start} &
\colhead{Exp.\ Time (s)} &
\colhead{S/N per resel\tablenotemark{1}}
}
\startdata
13363 & COS  & G160M & 1611 & 1420--1594, 1612--1786 & 56658.74330 & 4232 & 49 at \ion{C}{4} $\lambda$1549 \\
16115 & COS  & G130M & 1291 & 1134--1274, 1291--1431 & 59181.06719 & 2038 &  9 at \ion{N}{5} $\lambda$1239 \\
16115 & COS  & G160M & 1611 & 1420--1594, 1612--1786 & 59181.12986 & 2088 & 21 at \ion{C}{4} $\lambda$1549 \\
16115 & STIS & G230L & 2376 & 1570--3180             & 59181.20053 & 1043 & 85 at \ion{Mg}{2} $\lambda$2800   \\
16115 & STIS & G430L & 4300 & 2900--5700             & 59181.21917 &  144 & 42 at 4000 \AA  \\
16115 & STIS & G750L & 7751 & 5240--10,270           & 59181.22334 &   32 & 33 at 5700 \AA \\
\enddata
\tablenotetext{1}{Signal to noise ratio per resolution element. The spectral resolving power is $\sim$ 16,000 for the COS spectra and $\sim$ 500--1000 for the STIS spectra.}
\end{deluxetable*}

\subsection{X-ray Data}
X-ray data will provide complementary diagnostics of accretion in soft X-rays and coronal activity as well as critical inputs for evaluating disk irradiation for photo\-evaporation and MHD effects. Dedicated X-ray programs have been secured for three of the four ULLYSES monitoring targets. Grating spectroscopy will be obtained simultaneously with {\it HST} for TW~Hya, BP~Tau, and RU~Lup by the X-ray Multi-Mirror Mission-Newton and Chandra. Also, the Neutron Star Interior Composition Explorer (on the International Space Station) will be scheduled simultaneously with FUV to provide charge-coupled device (CCD)-type spectroscopy for all four targets in the monitoring sample to obtain flux information. 
 
Complementary X-ray data will come from the eROSITA all-sky survey \citep{Predehl2021}. Most ULLYSES targets are expected to be eventually detected by eROSITA given the targeted X-ray flux limit of $10^{-14}\,\escm$ \citep{Merloni2012}, approximately equivalent to a young 0.1\msun\ star at 150\,pc.
 
\subsection{Future Data}  

The ODYSSEUS survey will also aim to obtain data with {\it JWST} to probe connections between the UV radiation field and infrared emission lines. We will also propose for data with ALMA to measure important disk properties such as disk masses and inclinations.

\section{Observations of CVSO~109} \label{obs}

The previous section describes our broad goals for ancillary observations. In this section, we describe the datasets we obtained for CVSO~109 to use in our analysis.  The system is a CTTS-WTTS binary separated by $0\farcs64$, with similar brightness at red wavelengths and blue emission dominated by the accreting component, CVSO 109A (see this section and \S 4).  Only observations with sub-arcsecond angular resolution ({\it HST} and Gaia) resolve the two components.  The two components are assumed to be a physical binary rather than a visual binary.

\subsection{HST}

CVSO~109 was observed with the {\it HST} on 2020 November 28 as part of ULLYSES program 16115 (PI Roman-Duval). All ULLYSES {\it HST} data for CVSO~109 were obtained within a 4~hour window. Observations occurred over two visits, one with the COS FUV channel and one with the STIS NUV and optical settings. The COS visit lasted for two orbits and obtained spectra at two settings, both with spectral resolving power $\sim16,000$. The first setting used the COS/G130M grating with a central wavelength of 1291~\AA. Two exposures were obtained, with the spectra offset from one another by $\sim250$ pixels (2.5~\AA) in the dispersion direction to reduce fixed-pattern noise in the final coadd. The second setting used the COS/G160M grating with a central wavelength of 1611~\AA. Four exposures were obtained, again each offset from the previous by $\sim250$ pixels (3.1~\AA) in the dispersion direction to reduce fixed-pattern noise. COS FUV spectra were generated in two disjointed segments with a central gap of width $\sim15$~\AA. We note that the observing strategy for the COS/G160M grating was refined following the Orion OB1 observations, using two central wavelength settings that fill each other's gaps.

The STIS visit lasted for one orbit and obtained spectra at three settings, all with spectral resolving power $\sim500$--1000. The first setting used the NUV Multi-Anode Micro-channel Array (MAMA) and the STIS/G230L grating, while the remaining two used the CCD and the STIS/G430L and STIS/G750L gratings. The STIS observations detected a subarcsecond companion \citep{proffitt21}. 

We include in the analysis a COS spectrum obtained on 2014 January~1 in program 13363 with the same COS/G160M setting described above. The wavelength coverage, time of observation, exposure time, and S/N ratio of each Hubble spectrum appear in Table~\ref{tab:hst}.

Spectra processed with the CalCOS and CalSTIS pipelines appear in the MAST archive,\footnote{https://mast.stsci.edu/} while separate calibrated spectra for each of the components visible in the STIS slit are available as special ULLYSES data products.\footnote{https://ullyses.stsci.edu/ullyses-download.html} Wavelength errors are dominated by zero-point uncertainties, expected to be $\pm0.03$--0.04~\AA\ for the COS medium-resolution spectra and $\pm1$--2~\AA\ for the STIS low-resolution spectra. Absolute flux uncertainties are $\sim5\%$.

\begin{table*}
\centering
\caption{ODYSSEUS Spectroscopic and Photometric Observations} \label{tab:otherobs}
\begin{tabular}{cccccc}
\hline
\multicolumn{6}{c}{Spectroscopic Observations}\\
Telescope & Instrument & MJD & Wavelength Range & Resolution & PI \\ 
\hline
McDonald 2.7 m & Tull Echelle$^{a}$ & 59164.49 &  3850--10240 \AA\ & 60,000 & Johns--Krull \\
VLT & UVES & 59179.14 & 3300--4500 \AA, 4800--6800 \AA\ & 70,000 & Manara \\
VLT & UVES & 59180.13 & 3300--4500 \AA, 4800--6800 \AA\ & 70,000 & Manara \\
SMARTS/CTIO 1.5m & CHIRON & 59180.68 & 4080--8900 \AA\ & 27,800 &  Walter$^{b}$ \\
VLT & X-shooter$^{c}$ & 59181.15 & 10000--25000 \AA\  & 11,600 & Manara \\
VLT & UVES & 59181.17 & 3300--4500 \AA, 4800--6800 \AA\ & 70,000 & Manara \\
SMARTS/CTIO 1.5m & CHIRON & 59181.70 & 4080--8900 \AA\ & 27,800 &  Walter$^{b}$ \\
SMARTS/CTIO 1.5m & CHIRON & 59182.73 & 4080--8900 \AA\ & 27,800 &  Walter$^{b}$ \\
\hline
\multicolumn{6}{c}{Photometric Observations}\\
Telescope & Instrument & MJD & Filters &  Program & PI  \\
\hline
Konkoly RC80 & FLI PL230 CCD & 59159.95 -- 59209.96 & $BV$$r'i'$  & \dots            & K\'osp\'al \\
AAVSOnet     & various       & 59163.69 -- 59202.85 & $BV$$r'i'$  & AAVSOnet/195     & Walter \\
AAVSO        & various       & 59168.75 -- 59265.61 & $BVRI$  & Alert Notice 725 & Walter \\
OACt 0.91-m  & KAF-1001E CCD & 59178.95 -- 59200.09 & $BVRIZH\alpha$ &        \dots          & Frasca \\
TESS  & FFI & 59174.23 -- 59200.23 & \dots &    \dots  & \dots  \\

\hline
\end{tabular}
\tablecomments{$^{a}$Echelle E2. $^{b}$Program 471. $^{c}$Here, we use the NIR data only.}
\end{table*}

\subsection{CHIRON}

We obtained spectra of CVSO~109 using the CHIRON bench-mounted, fiber-fed, cross-dispersed echelle spectrograph on the 1.5~m telescope at the Cerro Tololo Inter-American Observatory (CTIO) on the Small and Moderate Aperture Research Telescope System (SMARTS) \citep{tokovinin13}. The data were taken in ``fiber mode" with $4\times4$ on-chip binning yielding a resolution $\lambda/\delta\lambda\sim27,800$. Wavelength coverage is complete from 4080~\AA\ through 8262~\AA\ in 70 orders, with incomplete coverage to 8900~\AA\ due to interorder gaps between the last five orders.

CHIRON data of CVSO~109 were taken on 2020 November 27, 28, and 29 with UT start times of 04:19:57, 04:51:30, 05:29:24 and exposure times of 900~s, 900~s, 1200~s, respectively. Integration times were 15 to 20 minutes, in single integrations on three successive nights centered on the time of the {\it HST} observations. The COS spectrum preceded our second spectra, which took place during the STIS observation.

The data were reduced using a pipeline coded in IDL.\footnote{http://www.astro.sunysb.edu/fwalter/SMARTS/CHIRON/ch\_reduce.pdf} The images were flat-fielded. Cosmic rays are removed using the L.A.Cosmic algorithm \citep{vandokkum01}. The 75 echelle orders were extracted using a boxcar extraction, and instrumental background, computed on both sides of the spectral trace, was subtracted. Since CHIRON is fiber fed, there is no simple method to subtract the sky. The fiber has a diameter of 2.7~arcsec on the sky. In any event, for bright targets, night-sky emission is generally negligible apart from narrow [\ion{O}{1}] and Na~D lines and some OH airglow lines at longer wavelengths.

Wavelength calibration relies on thorium--argon calibration lamp exposures at the start and end of the night and occasionally throughout the night. CHIRON in fiber mode is stable to better than 250~m~s$^{-1}$ over the course of many nights. The instrumental response was removed from the individual orders by dividing by the spectra of a flux-standard star, $\mu$~Col. This provides flux-calibrated orders with a systemic uncertainty due to sky conditions. The individual orders are spliced together, resulting in a calibrated spectrum at 4080--8900~\AA. 

All wavelengths are corrected to heliocentric. Radial velocities (RVs) are measured by cross-correlating a line list appropriate for cool star photospheres (mostly \ion{Fe}{1} and \ion{Ca}{1}) with the 5000--7500~\AA\ spectrum. Uncertainties on the individual RVs are 2--3~$\kms$. Equivalent widths are measured above a continuum interpolated from the adjacent continuum. Details about the spectroscopic observations using CHIRON and other instruments are found in Table~\ref{tab:otherobs}.

\subsection{McDonald Observatory 2.7 m Spectra}

A single high-resolution spectrum ($R\sim60,000$) of CVSO~109 was obtained with the Robert G.\ Tull cross-dispersed echelle spectrometer \citep{tull1995} coupled to the McDonald Observatory 2.7 m Harlan J.\ Smith Telescope. A $1.2^{\prime\prime}$ slit was used to record the spectrum in 56 orders on a Tektronix $2080\times2048$ CCD. A 2700~s exposure was recorded on 2020 November 11 starting at UT 11:19:38. The spectrum covers the wavelength range from 3850 to 10240~\AA, with complete coverage shortward of 5600~\AA\ and small gaps longward. Before and after the stellar spectrum, a comparison thorium-argon lamp spectrum was taken to determine the wavelength scale for the observations. The final wavelength scale adopted is the average one determined from the two lamp exposures. The spectrum was reduced with a custom package of IDL echelle reduction routines based largely on the data reduction procedures described by \citet{valenti_photospheric_1994} and \citet{hinkle_phoenix_2000}. The reduction procedure is standard and includes bias subtraction, flat-fielding by a normalized flat spectrum, scattered light subtraction, and optimal extraction of the spectrum. The blaze function of the echelle spectrometer is removed to first order by dividing the observed stellar spectra by an extracted spectrum of the flat lamp. Final continuum normalization was accomplished by fitting a low-order polynomial to the blaze-corrected spectra in the regions around the lines of interest for this study. The wavelength solution for the two comparison lamp spectra was determined by fitting a two-dimensional polynomial to $n\lambda$ as a function of pixel and order number, $n$, for approximately 1,800 extracted thorium lines observed from the internal lamp assembly.

\subsection{VLT Spectra}

We also incorporate optical spectra from X-shooter and UVES as part of the PENELLOPE Large Program (see \citealt{manara21} for details). X-shooter provides medium-resolution ($R\sim10,000$--20,000) spectra from 3000 to 25000 \AA. UVES provides $R\sim70,000$ spectra over wavelengths of 3300--4500 \AA\ and 4800--6800 \AA. The X-shooter data were taken on 2020 November 28 at UT 03:36:03.569. The UVES data were taken on 2020 November 26, 27, and 28 at UT 03:29:45.199, 03:02:45.220, and 03:59:35.437. 

\subsection{TESS}

\begin{figure}
\includegraphics[width=\columnwidth,angle=0]{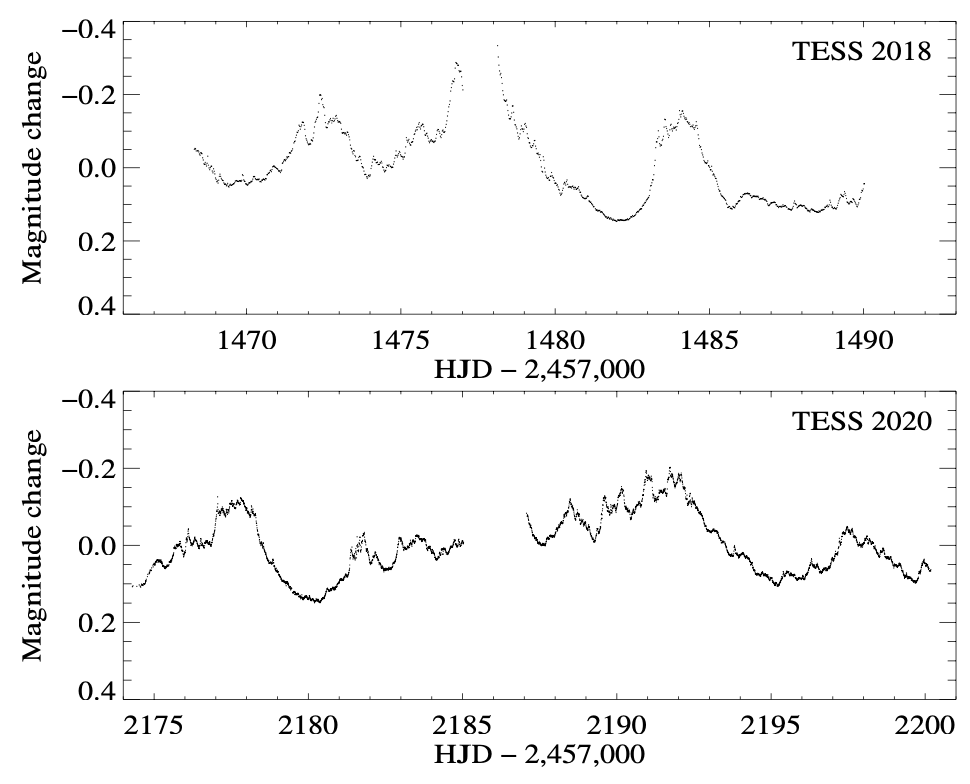}
\caption{{\it TESS} light curves of CVSO~109 from the 2018 (sector 6) and 2020 (sector 32) observing seasons, showing magnitude changes relative to the mean brightness in each season.}
\label{fig:TESS}
\end{figure}

CVSO~109 was observed by {\it TESS} \citep{Ricker14} in sector~32 (2020 November 19 to December 17), simultaneously with the other ULLYSES targets in the Orion OB1 star-forming region. It had also been observed two years earlier during {\it TESS} sector~6. We downloaded the full frame image data from the MAST archives using the TESScut software \citep{brasseur19}. {\it TESS} images, while photometrically stable and of continuous cadence, suffer from coarse spatial resolution (21\arcsec\ pixels) and a lack of color information. The {\it TESS} passband is 600--1000~nm. The temporal resolution is 10 minutes in 2020 and 30 minutes in 2018. We extracted the data in two independent ways using aperture photometry with 1.5 and 2.1 pixel radii, respectively. In the first case, the background was extracted from an annulus between 5 and 10 pixels from the source; in the second, the background was set to the mode of the counts within a $35\times35$ pixel box centered on the target, after excising all bright sources. The resulting light curves are nearly identical. To check if any other source fell in the {\it TESS} aperture and possibly contaminated the photometry, we downloaded a list of sources within $1'$ of the two components of CVSO~109 from the Gaia Early Data Release~3 (EDR3) catalog \citep{gaia21}. Gaia resolved CVSO~109, with G\,$=13.795\pm0.013$\,mag for 109A, G\,$=14.473\pm0.008$\,mag for 109B, and a separation of 0.636\arcsec. All the remaining stars in the $1'$ vicinity of CVSO~109 are $>3.0$\,mag fainter than the primary component and $>16$\arcsec\ farther away from it. The relative errors of the Gaia fluxes inform us about possible variability. We compared these uncertainties with the corresponding uncertainties of other stars of the same brightness using Figure~5.15 from the Gaia EDR3 Documentation V1.1\footnote{gea.esac.esa.int/archive/documentation/GEDR3/}. This comparison suggests that both CVSO~109A and CVSO 109B are variable stars.  All other, fainter stars in their vicinity are consistent with having a constant brightness. This suggests that the variability signal seen in the {\it TESS} data can be attributed to the CVSO~109 system, and contamination from other sources is negligible. The {\it TESS} light curve is shown in Figure~\ref{fig:TESS} and discussed further in Section~\ref{sec:variability}.

\subsection{Ground-based Optical Photometry}

To characterize the temporal behavior of CVSO~109 around the epoch of the {\it HST} spectra, optical photometric monitoring was carried out at different observatories. Here we combine data obtained with the RC80 telescope at the Piszk\'estet\H{o} Montain Station of Konkoly Observatory between 2020 November 6 and December 27, the AAVSOnet network of robotic telescopes between 2020 November 10 and December 19, and the M.~G.~Fracastoro station of the Osservatorio Astrofisico di Catania (OACt) between 2020 November 25 and December 17. During the Konkoly and AAVSOnet observations, images were obtained with Bessel $BV$ and Sloan $r'i'$ filters, while at OACt, broadband Bessel $BVRIZ$ and narrowband $H_{\alpha9}$ and $H_{\alpha18}$ filters were used. The narrowband filters have full-widths at half-maximum (FWHM) of 9 and 18\,nm, respectively. For the Konkoly observations, exposure times per image were 10, 20, 60, and 90~s in $BVr'i'$, respectively. Photometry was obtained using point-spread-function fitting, and the photometric calibration was done using APASS9 magnitudes \citep{henden15} of $\sim100$ stars within $12'$ of the target. For the AAVSOnet observations, exposure times per image were typically 5~minutes in $B$ and 3~minutes in $Vr'i'$. Here, aperture photometry was extracted, while the photometric calibration was done using APASS10 magnitudes of a few hundred stars within about $40'$ of the science target. For the photometric calibration of the OACt data, we refer to \citet{manara21} and \citet{frasca18}. Photometric values taken with the same filter on the same night were averaged. We supplemented these data with $g$-band photometry from the All Sky Automated Survey for Supernovae (ASAS-SN) survey \citep{shappee14,kochanek17}, where we also calculated nightly averages. When comparing the Konkoly and AAVSOnet photometry, we found small systematic offsets of 0.056, --0.005, 0.025, and 0.082~mag in $BVr'i'$, respectively. These are probably due to differences in the telescopes and filters, set of comparison stars, and extraction methods. To obtain a consistent dataset, we applied these small shifts to the AAVSOnet values. To convert the OACt Bessel $RI$ photometry to Sloan $r'i'$ values, we used conversion formulae from \citet{jordi06}. Here again, we needed small shifts between the OACt and Konkoly photometry, which were --0.010, --0.103, --0.127, --0.026\,mag in $BVr'i'$, respectively. The light curves are discussed further in Section~\ref{sec:variability}.

\begin{deluxetable*}{ccccccccccc}  
\tablecaption{Stellar Parameters of CVSO~109 \label{tab:starparam}}
\tablehead{
 \colhead{Star} & \colhead{SpT} & \colhead{$T_{\rm eff}$} & \colhead{$A_V$} & \colhead{$M_\star$} & \colhead{Log($L_\star$)} & \colhead{$R_\star$} & \colhead{J} & \colhead{H} & \colhead{K$_s$} & \colhead{EW(\halpha)}\\
  \colhead{} & \colhead{} & \colhead{(K)} & \colhead{(mag)} & \colhead{($M_{\sun}$)} & \colhead{} & \colhead{($R_{\sun}$)} & \colhead{(mag)} & \colhead{(mag)} & \colhead{(mag)} & \colhead{(\AA)}
}
\startdata
A & M0$_{-0.5}^{+0.5}$ & 3767.6$_{-81.2}^{+81.2}$ & 0.06$_{-0.24}^{+0.24}$ & 0.50$_{-0.05}^{+0.07}$ & --0.226$_{-0.110}^{+0.110}$ & 1.81$_{-0.25}^{+0.25}$ & 11.62$_{-0.11}^{+0.11}$ & 10.98$_{-0.13}^{+0.13}$ & 10.33$_{-0.16}^{+0.16}$ & --50$_{-1}^{+1}$ \\
B & M1$_{-0.5}^{+0.5}$ & 3639.4$_{-63.4}^{+63.4}$ & 0.55$_{-0.29}^{+0.29}$ & 0.42$_{-0.04}^{+0.04}$ & --0.314$_{-0.115}^{+0.115}$ & 1.75$_{-0.24}^{+0.24}$ & 11.91$_{-0.12}^{+0.12}$ & 10.99$_{-0.13}^{+0.13}$ & 10.74$_{-0.21}^{+0.21}$ & $<$--2.0
\enddata
\tablecomments{We adopt an RV of the CVSO~109 system of $16.51\pm0.50~\kms$\ and assume a distance of 400~pc from \citet{briceno19}. See Section~\ref{sec:stellarproperties} for details on the derivation of RV, SpT, $T_{\rm eff}$, $A_V$, $M_\star$, $L_\star$, the JHK photometry, and EW(\halpha). EW(\halpha) is from the HST G750L STIS spectrum; it varies with time, as discussed below. $R_\star$ is calculated from $L_\star$ and $T_{\rm eff}$. Since our derived $A_V$ is $\sim0$, we do not deredden the datasets in this paper in order to avoid introducing uncertainties from our choice of reddening law. 
}
\end{deluxetable*} 

\subsection{eROSITA X-ray Data}
eROSITA scanned over CVSO~109 during the eRASS1 all-sky survey between 2020 March 21st to 27th, detecting a total of $29\pm1.7$ counts about 7\arcsec\ from the nominal position of CVSO~109. Since we expect CVSO~109 to be an X-ray source and also assume the absence of any other known background source, we attribute this eROSITA source to CVSO~109. Fitting the X-ray spectrum  with an absorbed thermal plasma emission model results in $\log F_X=-12.63^{-0.53}_{+0.33}$ ($\escm$) within the 0.2--1.0\,keV band. For the assumed distance of 400\,pc, we arrive at an X-ray luminosity for both components of $3.0\times10^{30}$\,erg\,s$^{-1}$, in agreement with expectations for the sum of two young half-solar mass stars.

\section{Star and Disk Properties of CVSO~109} \label{sec:stellarproperties}

Here we show analysis and results for one object in the ULLYSES sample in order to preview some of the work that will be undertaken by ODYSSEUS. CVSO~109, a CTTS in the $\sim5$~Myr old Orion OB1b subassociation \citep{briceno19}, is known to be an accreting object \citep{ingleby14,manara21} surrounded by a pretransitional disk (i.e., a disk with an inner disk separated from an outer disk by a large gap) with a gap of $\sim19$~au \citep{mauco18}. CVSO~109 was revealed to be a binary system with a separation of 0.7$\pm$0.1{\arcsec} by \citet{tokovinin20}. The ULLYSES {\it HST} observations \citep{proffitt21} resolved the system (with a separation of 0.64{\arcsec}) and found that the faint companion had no significant  accretion activity (i.e., there is weak \ion{Mg}{2} $\lambda$2800 emission and weak or negligible H$\alpha$ emission, see Table~\ref{tab:starparam}). The Gaia~EDR3 release \citep{gaia21} measured a binary separation for CVSO~109 of 0.635{\arcsec}, which corresponds to a projected separation of 254~au (see Section~\ref{sec:distance}). 

Here we determine the stellar properties of the two binary components (Table~\ref{tab:starparam}) by fitting their NUV--NIR continuum as originating from a photosphere and accretion shock on the stellar surface. Due to the lack of accretion activity from CVSO~109B, we assume all accretion and ejection activity in the system arises from CVSO~109A. We note that it is not clear if A and B are physical binaries, but given the separation of the components, the MIR and millimeter emission seen in CVSO~109 \citep{mauco18} is likely coming from a circumstellar disk around CVSO~109A. The centroid of emission from the Wide-field Infrared Survey Explorer \citep{cutri13} is also consistent with CVSO~109A. Given that $A_V\sim0$ is within the measurement uncertainties (Table~\ref{tab:starparam}), we do not deredden the datasets used in our analysis in order to avoid introducing uncertainties from our choice of reddening law. 
\subsection{Distance} \label{sec:distance}
We adopt the mean distance of the Orion OB1b region of 400~pc \citep{briceno19} as the distance to CVSO~109, following \citet{manara21}. The Gaia EDR3 parallax would lead to a distance of 418~pc; however, the high uncertainty and the large value for the renormalized unit weight error of 24.94 both indicate that the parallax measurement is unreliable \citep{gaia21}. \citet{kounkel18} did not assign a group for CVSO~109 in their clustering analysis of the Orion OB1b Association with Gaia DR2 astrometry. The closest stellar group, oriCC-1, is at a distance of 421.9~pc (parallax $=2.370\pm0.062$~mas). These distances are all roughly consistent with one another.

\subsection{SED of CVSO~109A}

\begin{figure}    
\epsscale{1.15}
\plotone{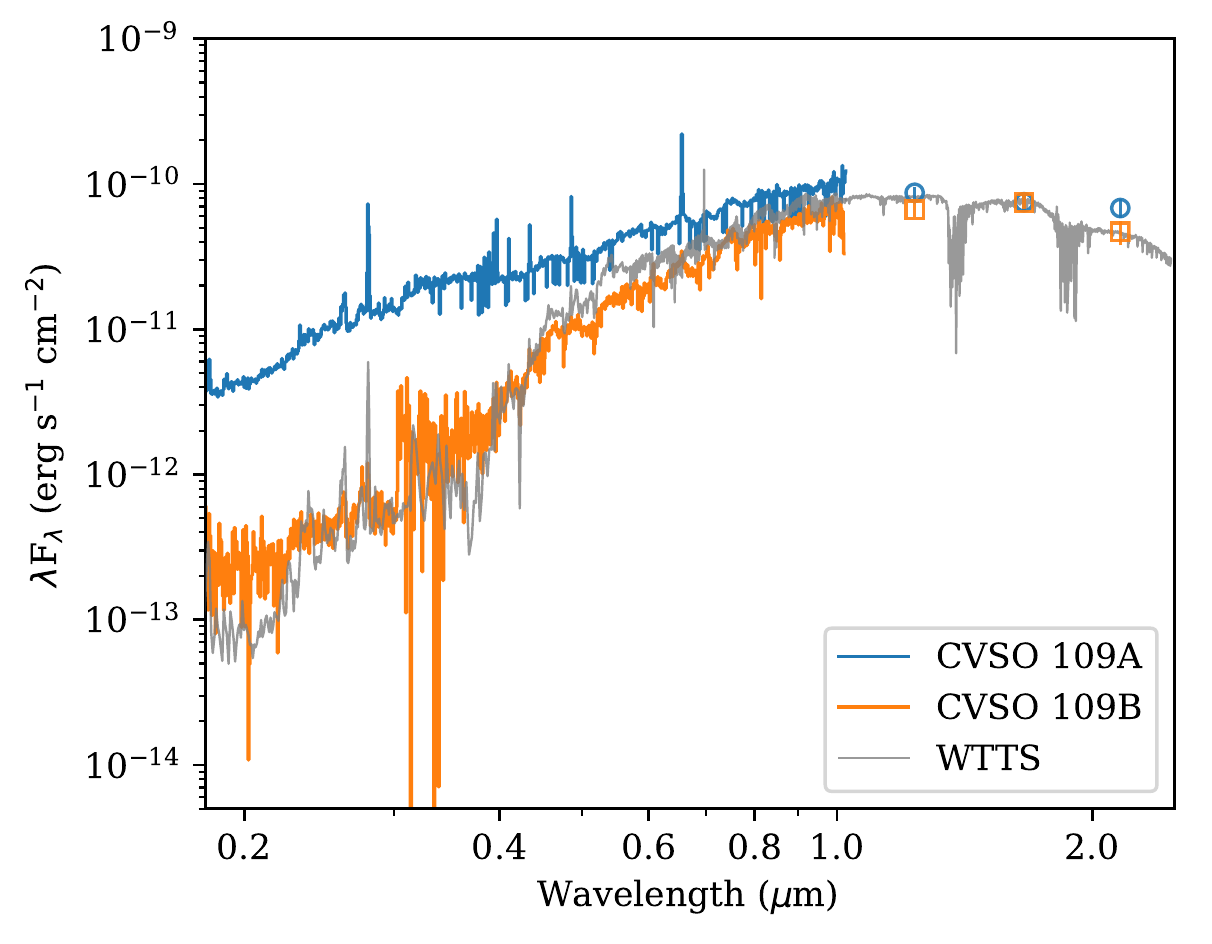}
\caption{
SEDs of CVSO~109A and CVSO~109B. We show {\it HST}/STIS spectra and NIR photometry derived from X-shooter and magnitude ratios from \citet{tokovinin20}. For comparison, we plot a combined STIS+X-Shooter WTTS spectrum used in Section~\ref{sec:accretionshock}. The spectrum is smoothed for clarity.
}
\label{fig:sed_binary}
\end{figure}

We construct individual spectral energy distributions (SEDs) for the A and B components from the resolved {\it HST} spectra for each of the CVSO~109 components from ULLYSES. Using the magnitude difference $\Delta m=m_B-m_A$ between CVSO~109A and CVSO~109B in resolved images from \citet{tokovinin20} and interpolating the flux-calibrated X-shooter spectrum to obtain J, H, and K$_{s}$ unresolved magnitudes, $m_{A+B}$, we can calculate the magnitudes of each component as
\begin{eqnarray}
    m_B &= m_{A+B}+2.5\log\left(1+10^{\Delta m/2.5}\right) ,  \\
    m_A &= m_B-\Delta m .
\end{eqnarray}
With the three magnitude differences provided by \citet{tokovinin20} for the three NIR bands, we can then produce NIR SEDs for the two components (Figure~\ref{fig:sed_binary}). The NIR magnitudes of each component are listed in Table~\ref{tab:starparam}.

In Figure~\ref{fig:sed_binary}, we also show the SED of a non-accreting star template, constructed from two WTTS as described in Section~\ref{sec:accretionshock}. CVSO109 B does not show any significant excess in the UV or near-IR, whereas the strong continuum excess is ubiquitous in the A component.

\subsection{Spectral Type and Equivalent Widths of H$\alpha$}

Using the {\it HST}/STIS spectra and the SpTClass tool \citep{Hernandez2017}, we obtain spectral types (SpT) and equivalent widths of \halpha\ for each component (Table~\ref{tab:starparam}). The effective temperatures were obtained using the conversion table of \citet{pecaut13}. 

We separately used our high-resolution spectrum from McDonald Observatory to constrain the effective temperature. The McDonald (and other ground-based) spectra do not separate the two components. For spectral type, we used the TiO band strength indices and relations defined by \citet{reid1995} and \citet{hawley1996}. In particular, we measured the TiO2 index, used \citet{hawley1996}'s methods to translate to the TiO5 index, and then used the relation of TiO5 index to spectral type reported in \citet{reid1995} to find a value of M0.5 to M0.8, depending on whether we use the dMe or dM TiO index relations. These resulting spectral types are between the values determined individually from the {\it HST} spectra, which makes sense since we are analyzing the combined light of the two stars. Since CVSO~109A dominates the light at the TiO2 wavelength (7043--7061~\AA, see Figure~2), we would expect the resulting spectral type to be closer to M0 than M1 for the nominal spectral types assigned in Table~3; however, given the spectral type uncertainty reported there and in the relation in \citet{reid1995}, our analysis of the high-resolution McDonald spectra is fully consistent with the resolved spectral types determined from the {\it HST}/STIS spectra measured above and from the unresolved UVES spectra measured by \citet{manara21}. 

\subsection{Extinction} \label{subsec:extinction}

Visual extinctions were obtained by comparing the optical Gaia~EDR3 and the near-IR J,H photometry with the standard colors for the respective spectral types \citep{Luhman2020}. For this estimation, we adopted the reddening law from \citet{Fitzpatrick2019}, assuming a canonical interstellar reddening law ($R_V=3.1$). We use a Monte Carlo (MC) error propagation method for estimating the visual extinction uncertainties reported in Table~\ref{tab:starparam}. Considering the observed error bars, we randomly vary the spectral types and the Gaia/J,H photometric colors to produce a distribution of possible results of visual extinction. We note that this approach leads to a range of uncertainties that includes negative extinction values. Negative extinction values are not physical but are the statistical results from the MC method. Here we provide the full range of uncertainty on the extinction estimate rather than impose restrictions to arrive at nonnegative extinction values. We note that our derived extinction does not consider a contribution from accretion but is consistent with the $A_V$ of 0.1 measured from fitting the combination of photospheric and accretion emission to the (unresolved) optical spectrum  \citep{manara21}. The extinction of CVSO~109B is larger than the extinction of the main component, which has an accreting disk (see Section~\ref{sec:accretion}). On the other hand, the small equivalent width (EW) of H$\alpha$ suggests that CVSO~109B does not have an accreting disk. If the CVSO~109 system is a physical binary with a projected separation of 254~au (Section~\ref{sec:stellarproperties}), the difference in extinction, if real, could be explained if the B component is behind the disk of the A component.

\subsection{Stellar Luminosity and Mass}

We use the bolometric correction from \citet{pecaut13} for the J-band and the adopted distance from \citet{briceno19} to estimate stellar luminosity for each component. Stellar masses of 0.50 and 0.42 \msun for CVSO 109A and CVSO 109B are estimated by comparing the stellar luminosities and the effective temperatures with the MIST evolutionary models \citep{Dotter2016}. For comparison, the masses obtained using the \citet{feiden16} magnetic tracks are $1.05_{-0.11}^{+0.10}$ and $0.82_{-0.10}^{+0.12}$ {\msun} for CVSO~109A and CVSO~109B, two times higher than the adopted masses. Here we also adopt the MC error propagation method for estimating uncertainties for the stellar luminosities and stellar masses. 

\subsection{Radial Velocity}

We used the McDonald spectrum to estimate the \vsini\ values and radial velocity (RV) of CVSO~109, again using the spectral range around the TiO bandhead at 7088~\AA, for comparison with the results obtained from VLT/ESPRESSO by \citet{manara21}. We used the spectrum at 7077--7104~\AA\ and compared the observation to corrected synthetic spectra of TiO from \citet{valenti1998}, computed for an effective temperature of 3700~K. We performed the comparison by convolving the synthetic spectrum computed at very high resolution with a Gaussian corresponding to a spectral resolution of $R=60,000$. We then binned the convolved synthetic spectrum onto the observed wavelength scale and used a cross-correlation analysis to measure the RV, fitting the peak of the cross-correlation function with a Gaussian and a parabola to get two estimates of the peak location, which were averaged. We repeated this multiple times, computing the synthetic spectra at different \vsini\ values starting at $2.0~\kms$\ and stepping up by $0.5~\kms$\ in each new iteration. We follow
\citet{nofi21}, basing their methodology on results summarized by \citet{gray08}, and assume a macroturbulent broadening of $2.0~\kms$\ when computing the synthetic spectra. We take the \vsini\ value $3.5\pm1.0~\kms$, which produces the strongest peak in the cross-correlation function, as the true value. The resulting RV we determine is $16.51\pm0.50~\kms$. The RVs and \vsini\ measured here are consistent with the 16--$17~\kms$\ RV and 3.2--$3.5~\kms$\ rotational velocity measured by \citet{manara21}. Both the
McDonald spectra analyzed here and that analyzed by \citet{manara21} do not 
resolve the two components of the system. Component A appears to dominate the light at the wavelengths
used (Figure 2), so these values are most appropriate for this component, which
is the accreting one of most interest to the current study.

\begin{figure*}
\includegraphics[width=\textwidth]{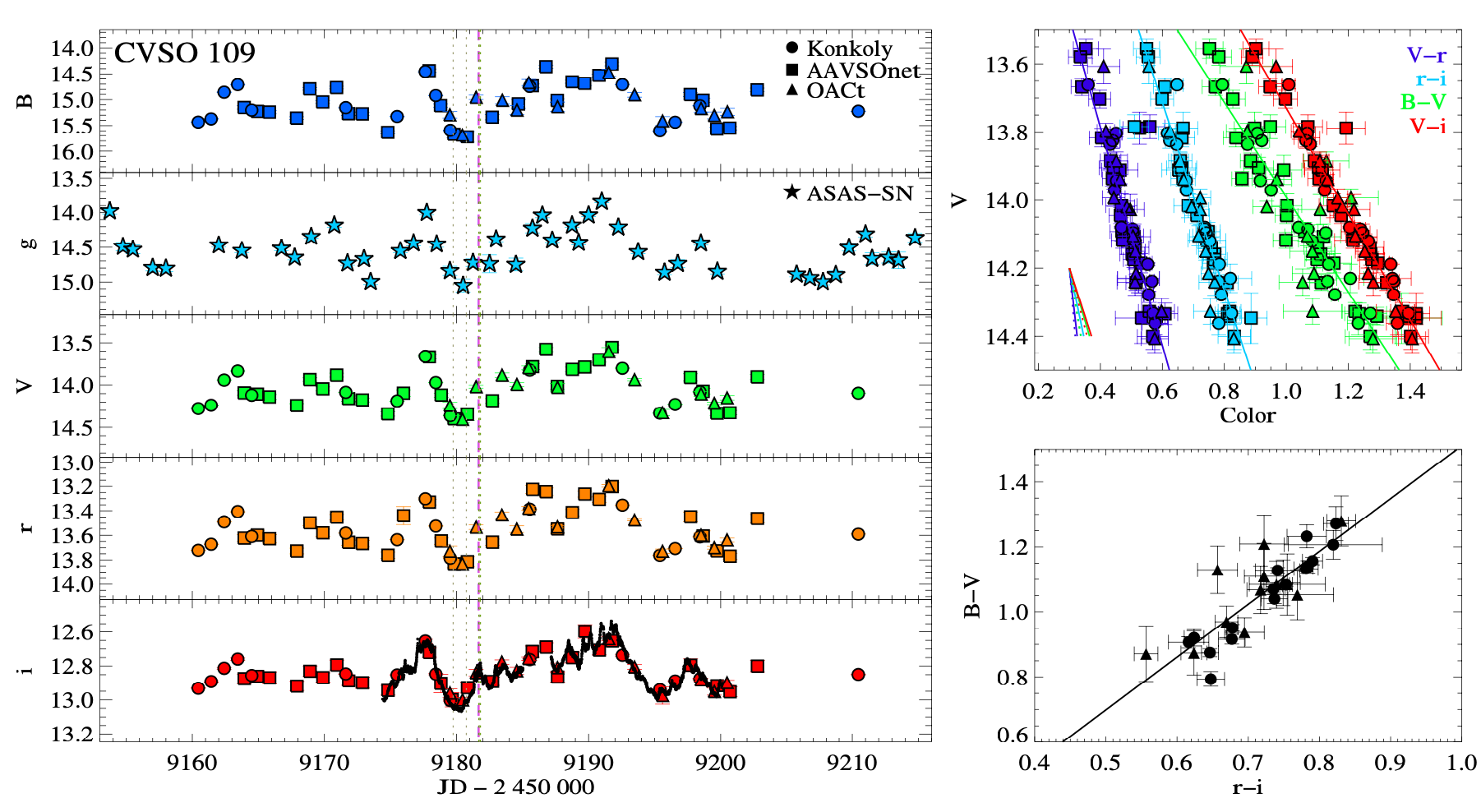}
\caption{Left: Optical light curves of CVSO~109. Data from Konkoly Observatory, AAVSOnet, OACt, and ASAS-SN are denoted with circles, squares, triangles, and stars. Vertical dashed and dotted lines mark the epochs of {\it HST}  and VLT observations.  In the bottom panel, the black curve displays the {\it TESS} data. Upper right: Color--magnitude diagram for CVSO~109. Lower right: Color--color diagram for CVSO~109. In the right panels, lines fitted to the data points are also plotted. The short lines in the upper right panel indicate how an increase of 0.2 mag in $A_V$ would change the colors for R=3.1 (solid lines) or for R=5 (dotted lines).} 
\label{fig:light}
\end{figure*}

\subsection{Inclination}

Disk inclinations of CTTS are best assessed with resolved submillimeter imaging, which is not yet available for CVSO~109.  Unresolved SED models point to a disk inclination of 53$^\circ$ \citep{mauco18}, but this approach is unreliable. We thus estimate the stellar inclination using the combination of {\vsini}, rotational period and stellar radius. The \vsini\ of the stellar photosphere is $3.5\pm1.0~\kms$, much lower than the rotational velocity measured for most young stars of similar spectral type \citep[see, e.g.][]{nguyen12}, already suggesting a near pole-on orientation. A Lomb--Scargle periodogram fit of a sinusoidal function to the {\it TESS} light curve suggests a periodicity of 6.5~days, using a variability analysis from the TESSextractor tool (Serna et al., submitted). Although the {\it TESS} light curve is a combined signal from both components, the accreting object CVSO~109A is 1.8 times brighter than CVSO~109B in this band and is expected to dominate the type of accretion bursts seen in the light curve. The $\sim6.5$ day period is recovered for both the 2018 and 2020 epochs from {\it TESS}, although quasiperiodicity may apply to the accretion flow rather than stellar rotation (or perhaps both) and may not be regular. 

Assuming a rotational period of 6.5~days, a \vsini\ of $3.5~\kms$, and a stellar radius of 1.8~$R_\odot$ (see Table~3), we estimate a relatively small rotational inclination angle ($\sim14^\circ$) for the star, indicating that the disk would be seen close to face-on if the star and disk are aligned.  

\section{Variability and the Epoch of {\it HST} Observations} \label{sec:variability}

One epoch of {\it HST} observations occurred during a local maximum in the light curve (Figure~\ref{fig:light}). Contemporaneous data provide further context for the interpretation of the state of CVSO~109 at the epoch of the {\it HST} observations.  In this section, we describe photometric and spectroscopic monitoring to place the UV observations in a broader context of accretion activity.

\subsection{Light Curves of CVSO~109}

Optical light curves of CVSO~109 are shown in Figure~\ref{fig:light} (left). None of these observations resolves the A and B components (see Section~\ref{sec:stellarproperties}), and therefore the light curves represent the total brightness of the two stars.

CVSO~109 displayed significant brightness changes, with multiple peaks and valleys during our observations. The peak-to-peak photometric variability amplitudes are $\Delta B=1.41$~mag, $\Delta g=1.21$~mag, $\Delta V=0.85$~mag, $\Delta r'=0.63$~mag, and $\Delta i'=0.41$~mag. The shape of the light curves is similar in all filters, but the amplitude of the variations decreases with increasing wavelength. $B$-band variations are 1.67 times larger than $V$-band variations, while $r'$- and $i'$-band variations are only 0.72 and 0.40 times those in the $V$-band, respectively. The tight correlations between the filters are demonstrated in the color--magnitude and color--color diagrams in Figure~\ref{fig:light} (right), where we fitted the data points with linear relationships. The slopes of the fitted lines are significantly shallower than what would be expected if the variations were due to changing interstellar extinction toward the star. The wavelength dependence of the variations is much steeper than that of the interstellar reddening.

In Figure~\ref{fig:light} we also compare our ground-based $i'$-band photometry with the much denser cadence {\it TESS} sector~32 light curve. The peak-to-peak variation during the 27~days of {\it TESS} observations is 0.33~mag. We note a very good correspondence between the two datasets, as expected for photometry obtained with filters that have a significant overlap between the wavelength ranges they cover. The {\it TESS} light curve and the ground-based data also show that CVSO~109 was actually close to the peak of a local brightness maximum when the {\it HST} spectra were taken 
which we interpret in the following sections as an increase in accretion. Thanks to the similar light curve shapes in different filters, the {\it TESS} observations can be used to construct multifilter, high-cadence light curves by linear transformations (i.e., by scaling the variability amplitude and by shifting the average brightness level of the {\it TESS} data). With this method, we estimated the following ``synthetic'' magnitudes for the epoch of the {\it HST} observations: $B=15.03$~mag, $g=14.45$~mag, $V=14.01$~mag, $r'=13.53$~mag, and $i'=12.83$~mag. Comparing these magnitudes with the SEDs in Figure~\ref{fig:sed_binary} confirms that these values represent the sum of CVSO~109A and CVSO~109B.

\subsection{Spectra of CVSO~109} \label{sec:Halpha_spectra}

\begin{figure}[!t]
\includegraphics[width=0.5\textwidth,trim=28mm 33mm 8mm 27mm]{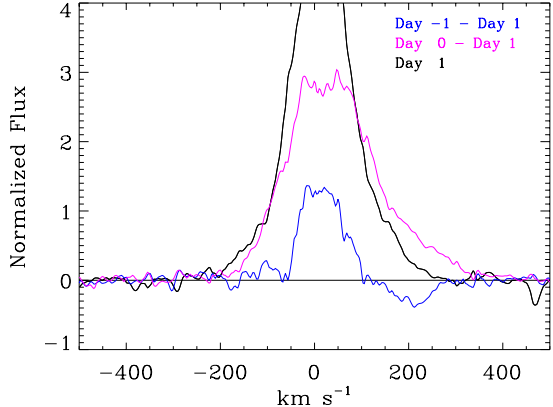}
\caption{An H$\alpha$ profile and two residual profiles from the 3 CHIRON spectra obtained on a daily cadence. These were obtained approximately 12 hours before, 12 hours after, and 36 hours after the HST spectra. We refer to these as days -1, 0, and +1 respectively. The black spectrum is the continuum-subtracted profile of H$\alpha$ on day~1, 1.5 days after the {\it HST} observations (day~1). This profile is symmetric; we use it as a reference. This line flux peaks at 6.3 in these units (7.3 continuum units). The other spectra are normalized to the continua and the reference spectrum is subtracted.
The residual spectrum taken before the {\it HST} observations (day~--1, in blue) shows a modest flux increase near line center and an inverse P~Cygni profile, suggestive of infalling gas. Twelve hours after the {\it HST} observations (day 0, magenta), the residual spectrum is brighter and broadened to the red. Figure~\ref{fig:halpha-color} shows that the H$\alpha$ emission was brighter during the UVES spectrum, obtained closer to the time of the {\it HST} spectrum. These velocities are heliocentric and are not corrected for the stellar radial velocity, which is insignificant on this scale.
}
\label{fig:hadiff}
\end{figure}

The three CHIRON spectra, obtained one day before (day~$-1$), during (day~0), and one day after (day~1) the {\it HST} observations (Figure~\ref{fig:hadiff}; see Table~\ref{tab:otherobs}), show striking line profile changes. The H$\alpha$ profile in the first CHIRON spectrum shows an inverse P-Cygni absorption feature at 200 $\kms$ that is particularly noticeable in the difference profile (Figure 4), suggestive of infalling material along the line of sight. This velocity is a fair fraction of the free-fall velocity and so likely arises in gas in free-fall near the bottom of an accretion funnel. The spectrum coincident with the STIS observation shows an enhanced red wing on H$\alpha$ emission, as well as an overall increase in line brightness. The final CHIRON spectrum, one day after the {\it HST} observations, shows a weaker and symmetric H$\alpha$ profile. In Figure~\ref{fig:hadiff}, we show the difference spectra, using the symmetric profile (day~1) as a reference. The variations in the red side of the line and in overall line strength are obvious, while the blue wing is only slightly enhanced.  Considering also the H-alpha profile observed at McDonald 16 days earlier, the overall impression is that while the line strength does vary, the shape of the blue wing stays fairly constant while there is substantially more variation in the shape of the red wing.

We interpret these results as indicating the passage of a discrete accretion flow across our line of sight on day~--1, with the impact on the star before the {\it HST} observations, resulting in the strong enhancement of the H$\alpha$ line flux on day~0. By day~1, the enhanced accretion event had ended. The {\it HST} observations likely occurred during a modest accretion event that had a few percent effect in the 600--1000~nm continuum brightness and drove a 60\% increase in the H$\alpha$ and H$\beta$ fluxes.

\begin{figure}
\includegraphics[width=8.5cm]{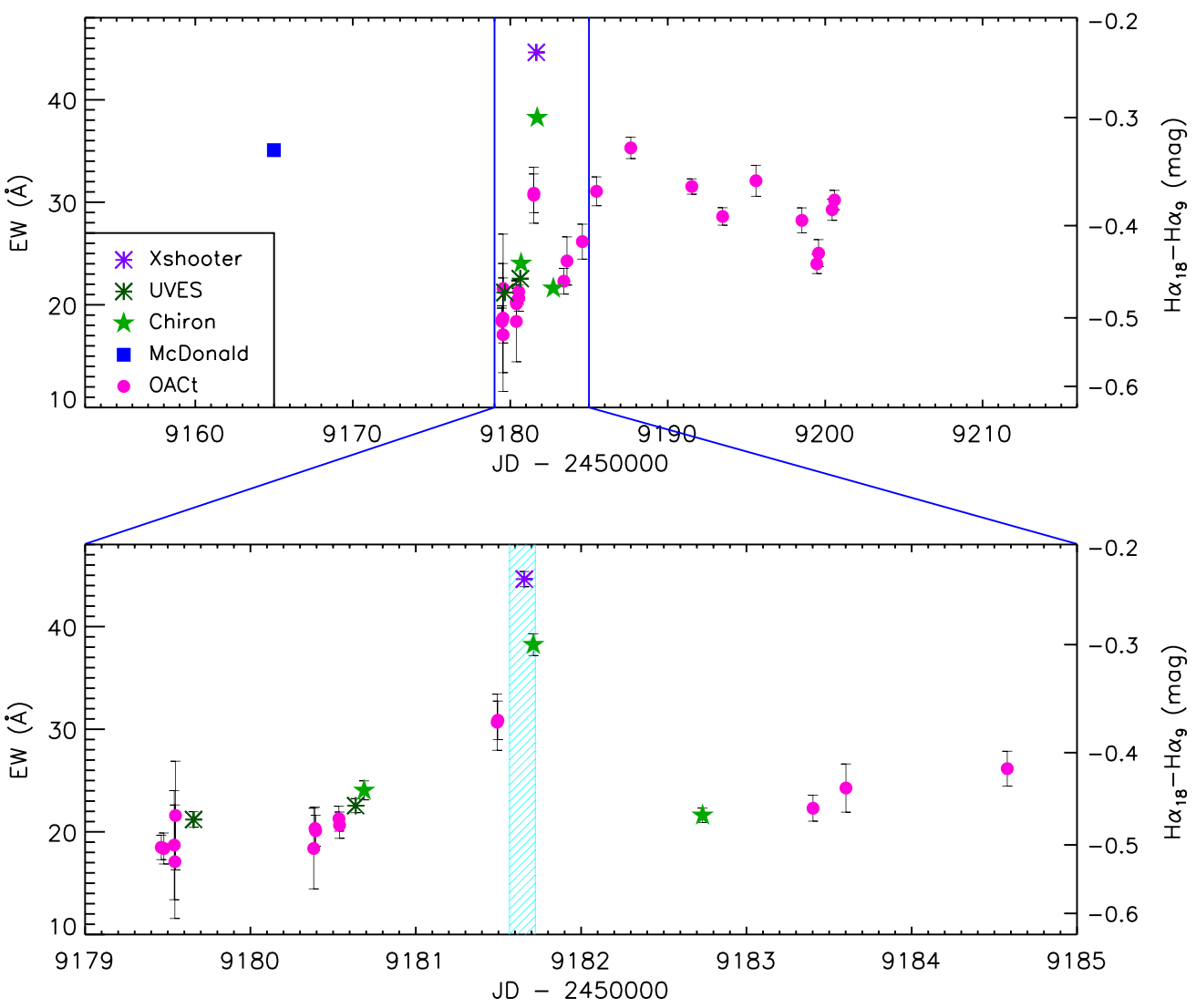}
\caption{H$\alpha$ EW derived from the color index measured at OACt (dots) over one month (top) and zoomed into a 15 day interval (bounded by the blue lines on the top plot) to highlight the epoch of {\it HST} observations. The scale of the color index is shown in the right vertical axis. The H$\alpha$ EW measured on X-shooter, UVES, CHIRON, and McDonald spectra are overplotted with different symbols as indicated in the legend. A zoom around the {\it HST} observations (cyan hatched area) is displayed in the lower box.}
\label{fig:halpha-color}
\end{figure}

To follow the temporal evolution of the H$\alpha$ intensity over a longer time scale, we used the narrowband photometry that was performed at the OACt for about a month, starting just before the {\it HST} observations. The color index $H\alpha_{18}-H\alpha_{9}$ is an indicator of line emission that can be converted into equivalent width by means of the calibrations of \citet{frasca18}. Since we have several high- or mid-resolution spectra of CVSO~109, we have used the ``synthetic'' color indices measured on the latter ones, by integrating them in the passbands of the filters, to calibrate the OACt color index into the EW. The H$\alpha$ EW is shown in Figure~\ref{fig:halpha-color} as a function of the Julian date. As can be seen, the EWs from the spectra are in excellent agreement with those measured from the OACt photometry and allow us to trace the time evolution of the H$\alpha$ intensification during {\it HST} observations, which appears to have started at least 5~hours before the time of the X-Shooter and the second CHIRON spectrum. Note also the considerable variation of the line intensity during the observing run, which roughly follows the broadband photometric variations.

\section{Accretion, Ejection, and Disk Irradiation}
\label{sec:accretion}

In this section, we derive the properties of the star-disk interactions, including accretion, winds, and disk irradiation, through measurements and models of the spectral lines and continuum emission. In Section \ref{sec:NUVtoNIR}, we fit the NUV--NIR continuum emission as originating from a combination of emission from the accretion shock on the stellar surface and emission from the irradiated inner edge of the dusty disk. In Sections \ref{sec:CIV_line} and \ref{sec:extendedflow}, we show preliminary accretion shock model fits of the FUV \ion{C}{4} line and fit CVSO~109's optical emission lines as arising from the accretion flow onto the star. Section \ref{sec:mdot_variability} discusses variability in the accretion rates reported here and in the literature. In Section \ref{sec:InnerDiskWind}, we study the wind of CVSO~109, showing FUV wind absorption lines and profiles and deriving the mass-loss rate. In Section \ref{sec:FUV_radiation_environment}, we investigate the FUV radiation environment of CVSO~109 by measuring the FUV continuum and reconstructing the Ly$\alpha$ profile; additionally, we derive the H$_2$ emitting region and analyze the UV-H$_{2}$ and \ion{O}{1} emission profiles. Lastly, in Section \ref{sec:FUV_variability} we discuss the FUV variability of CVSO~109.

\begin{figure*}    
\plotone{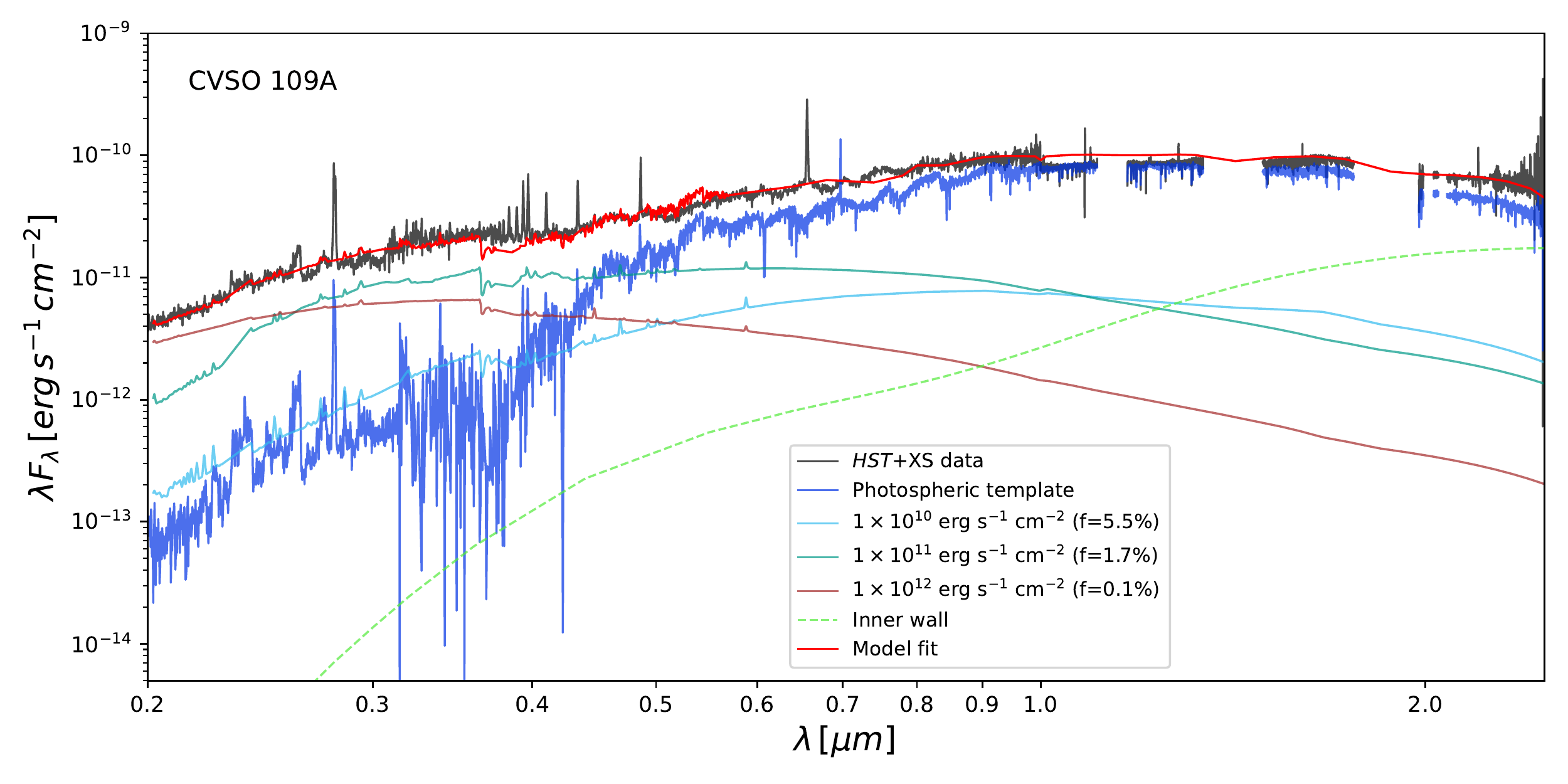}
\caption{Fitting the NUV--NIR continuum of CVSO~109A. We show the {\it HST} and X-shooter spectra (black) along with the photospheric template (dark blue), which is scaled using optical veiling derived by X-shooter. The parameters of the best-fitting model (red solid line) are listed in Table~\ref{tab:outmodelparam}. The best-fit model is a combination of the accretion shock model (consisting of the three columns with energy fluxes of $1\times10^{10}$, $1\times10^{11}$, and $1\times10^{12}$ $\escm$, each with filling factors shown in parentheses in the legend; cyan, sea-green, and brown solid lines, respectively) and the DIAD model (consisting of the emission from the inner wall; dashed line). Emission lines and regions strongly affected by telluric absorption were excluded from the fit. The gaps in the NIR spectra are the telluric regions, which are omitted from the plot to improve the readability of the figure.
}
\label{fig:NUVNIRcontinuum}
\end{figure*}

\subsection{NUV--NIR Continuum}
\label{sec:NUVtoNIR}

Here we fit the NUV--NIR continuum of CVSO~109A (Figure~\ref{fig:NUVNIRcontinuum}). The NUV and some of the optical emission arise from the accretion shock \citep{ingleby13}, while the NIR emission originates from the inner edge of the dust disk \citep{muzerolle03,espaillat10,mcclure13}. We use accretion shock models and disk models to reproduce the NUV--NIR continuum.

We assemble the continuum by combining {\it HST} NUV--optical spectra with an X-shooter NIR spectrum. The data were taken within about 3~hours of one another on 2020 November 28 and show little variability \citep{manara21}. We use the {\it HST} extracted spectrum of CVSO~109A and the resolved photometry discussed in the previous section to measure the flux ratio as a function of wavelength, and we use this to scale the unresolved X-shooter spectrum to the J-band flux expected from CVSO~109A. The spectra are shown in Figure~\ref{fig:NUVNIRcontinuum} along with the best-fitting models (discussed further in Section~\ref{sec:accretionshock}).

The contribution of CVSO~109B is small at short wavelengths, which justifies our approximation that the excess emission that produces the veiling in the unresolved X-shooter spectrum and the FUV emission in the COS aperture comes solely from CVSO~109A.  We adopt the X-shooter veiling at 550~nm \citep[$0.63\pm0.19$, ][]{manara21} to set the level of the photosphere. The photospheric template is constructed from two weak-line TTS (WTTS) stitched together. The template at 0.17--0.55~\micron\ is the {\it HST}/STIS spectrum of TWA~7 \citep[M1, ][]{webb99}, and the template beyond 0.55~\micron\ is the VLT/X-shooter spectrum of TWA~14 \citep[M0.5, ][]{manara13a}, which is the same template used in the analysis of CVSO~109 in \cite{manara21}.

\begin{deluxetable*}{ccccccccc}  
\tablecaption{Results of the Accretion Shock and Accretion Disk Models \label{tab:outmodelparam}}
\tablehead{
 \multicolumn{5}{c}{Accretion Shock Model} & \multicolumn{4}{c}{Accretion Disk Model} \\
 \cmidrule(lr){1-5}\cmidrule(lr){6-9}
 \colhead{$R_\star$ (\rsun)}  & \colhead{$\dot{M}$ ($10^{-8}$ \msunyr)} & \colhead{$f_{1E10}$} & \colhead{$f_{1E11}$} & \colhead{$f_{1E12}$} & \colhead{$a_{\rm max}$ ({\micron})} & \colhead{$z_{\rm wall}$ (H)} & \colhead{$T_{\rm wall}$ (K)} & \colhead{$R_{\rm wall}$ (au)}
}
\startdata
$1.80^{+0.13}_{-0.13}$ & $3.26^{+0.23}_{-0.24}$ & $0.055^{+0.012}_{-0.010}$ & $0.0172^{+0.0028}_{-0.0024}$ & $0.00091^{+0.00015}_{-0.00012}$ & 10 & 3 & 1400 & 0.09 \\ 
\enddata
\tablecomments{We use input parameters of $M_\star=0.50$ \msun, \ri=2.34~$R_\star$, T$_{\rm eff}=3767.6$~K, $A_V=0$, $r_{5500}=0.63\pm0.19$, and $d=400$~pc, and $i$ is taken to be $40^\circ$.}
\end{deluxetable*}

\subsubsection{Accretion Shock Modeling} \label{sec:accretionshock}

The SED at 0.17--0.57~\micron\ was fit using accretion shock models \citep{calvet98}. The accretion column is characterized by an energy flux ($\curf=1/2\rho v_s^3$), which measures the density of material in the accretion column ($\rho$), assuming that the magnetospheric disk truncation radius (\ri) and infall velocity ($v_s$, which depends on $R_\star$, $M_\star$, and \ri) are constant. Each column has a filling factor, $f$, which gives the fraction of the stellar surface covered by the column. 

Following \citet{ingleby13} and \citet{robinson19}, we use three accretion columns with $\curf$ of $10^{10}$, $10^{11}$, and $10^{12}$ $\escm$. The total energy flux ($\curf_{\rm tot}$) is the sum of the energy flux of all the regions weighted by their respective $f$. The total accretion column coverage on the stellar surface ($f_{\rm tot}$) is the sum of the $f$ values for all the regions.  

We adopt the measured values of $T_{\rm eff}$ and $M_\star$ (Table~\ref{tab:starparam}) as inputs to the model and assume $A_V=0$ as stated in Section \ref{sec:stellarproperties}. The derived value for $R_\star$ (Table~\ref{tab:starparam}) is taken to be a prior in the Markov chain Monte Carlo (MCMC) fit. The parameters for the best-fitting model in Figure~\ref{fig:NUVNIRcontinuum} are listed in Table~\ref{tab:outmodelparam}. Each value reported is the median of the posterior distribution for each parameter, and the associated uncertainties are the 16th and 84th percentiles (approximately 1$\sigma$ uncertainties for a Gaussian distribution). Additional uncertainty arises from ($i$) our choice to adopt $M_\star\sim0.50$ \msun\ from the \citet{Dotter2016} MIST models rather than $M_\star\sim1.05$ \msun\ from the \citet{feiden16} magnetic tracks, and ($ii$) our use of $A_V=0$. Using the upper limit of the \citet{feiden16} tracks ($M_\star=1.15$ \msun) would decrease the median accretion rate by a factor of 2.2. Including $A_V$ as a free parameter in the MCMC fit gives an upper limit of $A_V=0.39$, which would increase the median accretion rate by a factor of 1.6.

Our measured accretion rate for CVSO~109A of $3.26^{+0.23}_{-0.24}\times10^{-8}$ \msunyr\ corresponds to an accretion luminosity of $0.164^{+0.012}_{-0.012}$ \lsun. This is 28\% of the stellar luminosity reported in Table \ref{tab:starparam}.

\subsubsection{Accretion Disk Modeling}

Here we use the D'Alessio et al.\ Irradiated Accretion Disk (DIAD) models \citep{dalessio98,dalessio99,dalessio01,dalessio05,dalessio06} to fit the NIR continuum excess. The temperature and density structure of the disk are calculated iteratively, and the surface density of the disk is tied to the accretion rate. We adopt a pretransitional disk model, following the characterization by \citet{mauco18} that the CVSO 109 disk has a gap relative to a full disk \citep{espaillat10} with a frontally illuminated ``wall'' at the dust destruction radius along with another wall at the outer edge of the gap (i.e., at the inner edge of the outer disk). The inner wall at the dust destruction radius dominates the NIR and MIR emission (from $\sim5$--30~\micron), and the outer disk dominates the emission beyond $\sim40$~\micron. Therefore, here we include only the inner wall in our modeling. This has been shown to be appropriate in previous works fitting only NIR data \citep[e.g.,][]{espaillat10,mcclure13}.

The composition of the dust used in the disk model affects the resulting emission and derived disk properties \citep[see][]{espaillat10}.We include silicates and graphite with fractional abundances of 0.004 and 0.0025, respectively, following the \citet{draine84} model for the diffuse interstellar medium (ISM). We calculate the silicate (pyroxine) and graphite opacities using Mie theory and optical constants from \citet{dorschner95} and \citet{draine84}, respectively. The models assume spherical grains with a size distribution that scales as $a^{-p}$ between grain radii of $a_{\rm min}$ and $a_{\rm max}$ and $p$ of 3.5 \citep{mathis77}. $a_{\rm min}$ is fixed at 0.005~\micron\ while we vary $a_{\rm max}$ between 0.25~\micron\ and 10~\micron\ to achieve the best fit to the SED. Since $a_{\rm max}$  typically fits the silicate feature around 10~\micron, we note that the best-fitting value provided here is not well constrained and has been chosen to be 10~\micron\ in part to align with the best fit found by \cite{mauco18}, which fit the 10~\micron\ silicate feature of CVSO 109 well. 

The values for $z_{\rm wall}$ and $T_{\rm wall}$ are also adjusted to fit the SED. $z_{\rm wall}$ is the height of the wall, which we vary between one and five gas scale heights ($H$), and $T_{\rm wall}$ is the temperature at the surface of the optically thin wall atmosphere and ranges between 1200 and 1800~K. The radius in the disk at which the wall is located ($R_{\rm wall}$) is derived using the best-fitting $T_{\rm wall}$ following
\begin{equation}R_{\rm wall} \sim 
\left [{\frac{(L_* + L_{\rm acc})}{16 \pi \sigma_R} }
( 2 + { \frac{\kappa_s} {\kappa_d} }) \right ] ^ {1/2} 
{1 \over T_{\rm wall}^2 } 
\end{equation} 
from \citet{dalessio04}, where $\sigma_R$ is the Stefan-Boltzmann constant and $\kappa_s$ and $\kappa_d$ are the mean opacities to the incident and local radiation, respectively. $L_\star$ is the stellar luminosity and $L_{\rm acc}$ ($\sim GM_\star\dot{M}/R_\star$) is the luminosity of the stellar accretion shock. (For schematics of the inner disk wall and details of the temperature structure of the wall's atmosphere, see \citet{dalessio05}.) The values for $\dot{M}$ and $R_\star$ are taken from the best fit to the shock model described in Section~\ref{sec:accretionshock} to produce a self-consistent result. Parameters for the best-fitting DIAD model in Figure~\ref{fig:NUVNIRcontinuum} are listed in Table~\ref{tab:outmodelparam}. The best-fit parameters are generally consistent with those found by \cite{mauco18}, who used \mdot$=0.67\times10^{-8}$ \msunyr\ (as estimated from the H$\alpha$ line luminosity rather than the continuum excess); their best fit had z$_{\rm wall}=4$~H, $T_{\rm wall}=1400$~K, and $R_{\rm wall}=0.11$~au.

\begin{figure}    
\plotone{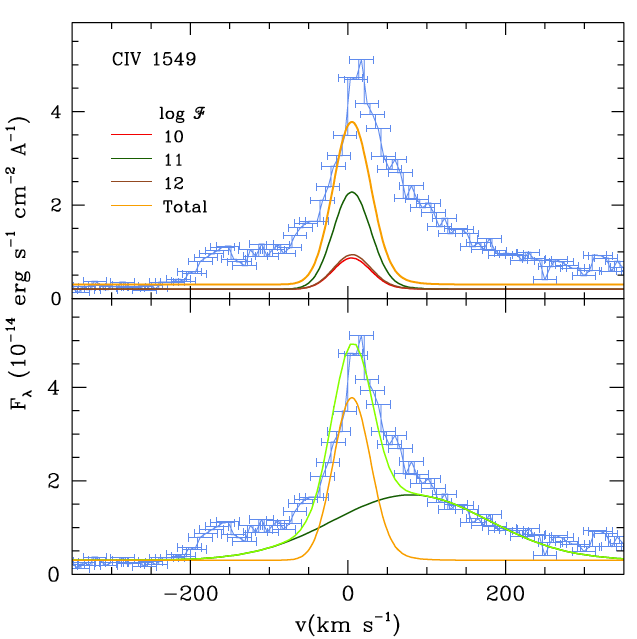}
\vskip -0.0in
\caption{
Profile of the {\civ} 1549 line in the CVSO~109A spectrum. The observed profile (blue) has been boxcar smoothed by 3 pixels. The upper panel shows the model line profile emitted by the postshock regions of columns with energy flux log $\curf=10$ (red), 11 (green), and 12 (brown), weighted by the filling factors that explain the NUV--NIR continuum (Figure~\ref{fig:NUVNIRcontinuum}). The total model line profile is shown in orange. The lower panel shows again the observed profile and the total postshock emission. A Gaussian with mean RV of $+80~\kms$\ and width 100~$\kms$\ (dark green) added to the total postshock emission can approximately explain the observed profile (light green).
}
\label{fig:civline}
\end{figure} 

\subsection{The FUV {\civ} Line in CVSO~109A}
\label{sec:CIV_line}

The observed line profile of the {\civ} 1549 line in CVSO~109A is shown in the upper panel of Figure~\ref{fig:civline}. The figure also shows the profiles predicted from the postshock regions of accretion columns with energy fluxes $\curf=10^{10}$, $10^{11}$, and $10^{12}~\escm$, each covering an emitting area $f4\pi R_\star^2$, with filling factors $f$ equal to those resulting from the fit of the NUV--NIR continuum (Figure~\ref{fig:NUVNIRcontinuum}, Table~\ref{tab:outmodelparam}). The total emission from the postshock region is the sum of the three columns.

To model the {\civ} 1549 line, we calculated the structure of the postshock region of each column, solving the fluid equations with boundary conditions given by the strong shock approximation \citep{calvet98,robinson19}. The emissivities were calculated using the Cloudy code \citep{ferland17}, which was also used to calculate the level populations of the \ion{C}{4} ion in the postshock region. We calculated the specific intensity of the line at each velocity in the line profile and at each inclination from the local normal solving the radiative transfer equation, with the source function and absorption coefficient calculated with the level populations; we assumed a Voigt profile and a turbulent velocity of $30~\kms$, to make the width consistent with the FWHM of the narrow component of the {\civ} 1549 lines in \citet{ardila13}. We assumed that the shock covered a ring on the stellar surface with a colatitude consistent with the extended flow modeling results (Section~\ref{sec:extendedflow}) and a total area $f\pi R_\star^2$. The contribution to the specific intensity of each velocity of the line profile from a given azimuthal angle along the ring depended on the RV of the point, which in turn depended on the colatitude of the ring, the inclination to the line of sight, and the azimuthal angle. The integral of the specific intensity over the ring at each velocity provided the line profile fluxes.

\begin{figure*}    
\epsscale{1.}
\plotone{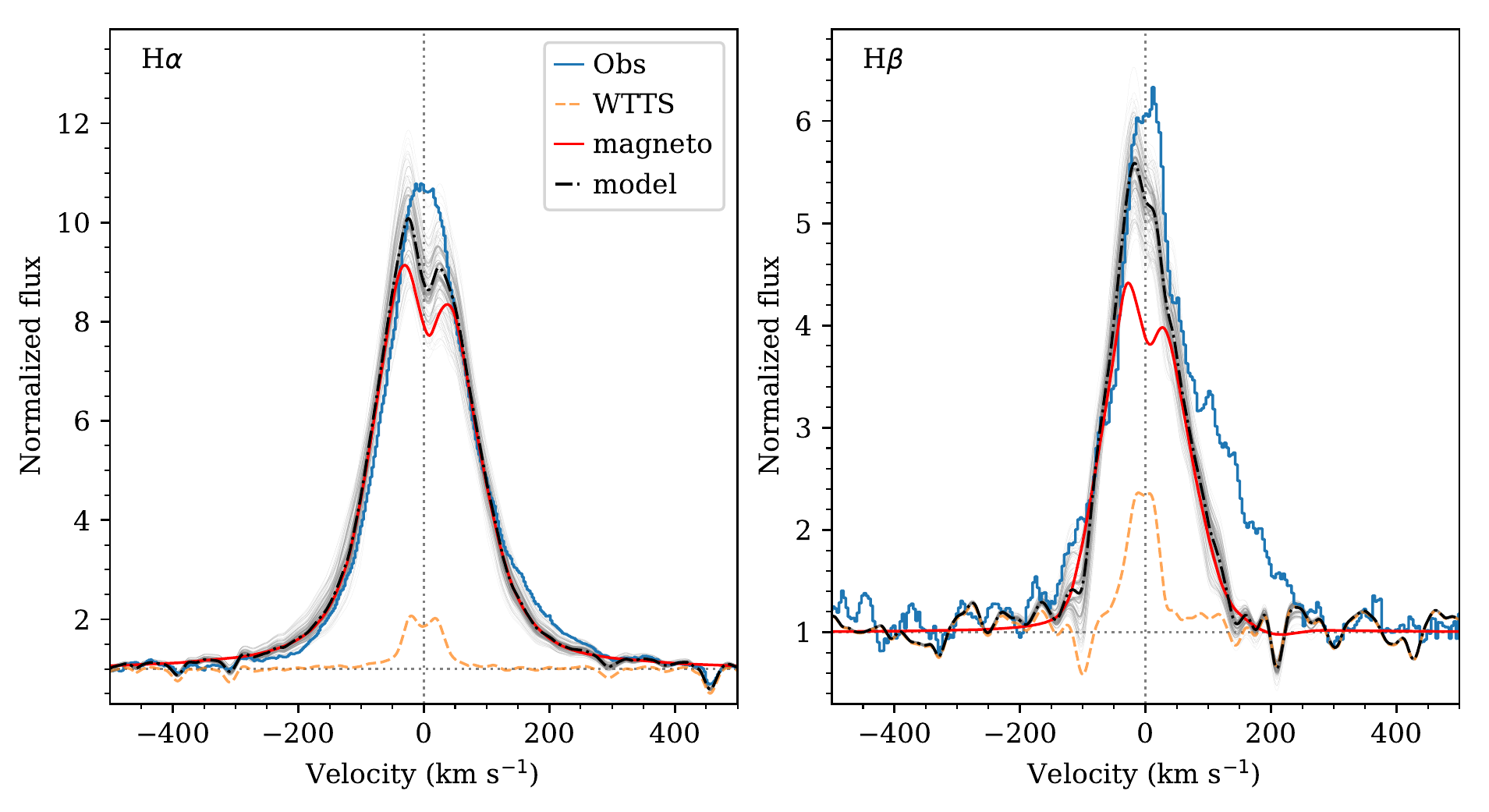}
\caption{
The extended accretion flow model fit to CHIRON spectrum of CVSO~109 in {\halpha} (left) and {\hbeta} (right) observed on MJD=59181.70. The observation is shown in blue. The template photosphere and chromosphere, shown in dashed orange lines, are from UVES observation of the M0 WTTS TWA~25. The light gray lines are 100 best fits, and the red lines and the dashed-dot black lines are the average magnetospheric profiles and the total magnetosphere + chromosphere + photosphere  model profiles, respectively.
}
\label{fig:accflow}
\end{figure*} 

The maximum of the predicted profile occurs at $8~\kms$, which is expected from the low velocities of the postshock region. For the mass and radius of CVSO~109A, the free-fall velocity for material falling from a disk truncation radius of $\sim2.4R_\star$, consistent with models of the extended region (Section~\ref{sec:extendedflow}), is 141~$\kms$; therefore, the maximum velocity in the postshock is 35~$\kms$, where the temperature is $\sim10^6$~K; the velocity at $T\sim10^5$~K, where the line forms, is even lower. The observed velocity profile peaks at $\sim10~\kms$, consistent with model predictions within the COS intrinsic velocity uncertainty of $\sim15~\kms$. 

Traditionally, line profiles have been fitted by the sum of Gaussian functions, adjusted to provide the best fit to the observed profile. In many cases, the line is decomposed into the narrow component (NC) and the broad component (BC) \citep{ardila13}. In the case of the UV metal lines, the NC has been identified with shock emission while the BC had been postulated to come from the magnetospheric flows, although only schematic  modeling has supported these statements so far \citep{calvet96}. The model results shown in Figure \ref{fig:civline} 
indicate that the emission from the postshock could explain the central peak of the line and could be assigned to the NC. A BC represented by a Gaussian with peak emission of $1.4\times10^{-14}~\escm$~\AA$^{-1}$, mean at $+80~\kms$ and width $100~\kms$ added to the postshock emission could explain the observed profile (Figure~\ref{fig:civline}). This BC could arise from hot regions in the magnetospheric flows that are not accounted for in the current models.

In this analysis, we have not included a contribution to the {\civ} 1549 line from the underlying chromosphere of the star. The line luminosity is $\sim10^{-4}$ \lsun, which is a factor of 10 or more higher than the line luminosity for most WTTS in the \citet{ardila13} sample, which we adopt as representative of the chromosphere. Given this, we do not expect a large chromospheric contribution to the observed highly ionized metal lines in this star, but we will readdress the issue in the future, more complete analysis of the FUV hot line spectrum.

\subsection{Models of the {\halpha} and {\hbeta} Emission Lines}
\label{sec:extendedflow}

The CHIRON spectra cover the {\halpha} and {\hbeta} lines, providing properties of the accretion flow via line profile modeling (see Figure~\ref{fig:accflow}). Here we model the {\halpha} and {\hbeta} lines in the spectrum observed during the {\it HST} epoch (day 0 in Figure~\ref{fig:hadiff}) since it is contemporaneous with the {\it HST} spectrum used for FUV line and NUV continuum modeling. These two lines were selected as they are the strongest magnetospheric lines and, therefore, they can be decomposed into different components more easily than weaker hydrogen lines. We use the magnetospheric accretion flow model of \citet{hartmann94,muzerolle98,muzerolle01}. Here we present our various model assumptions. The magnetic, stellar rotation, and disk rotation axes are aligned, and the material flows onto the star along an axisymmetric accretion flow arising from the corotating gas disk. A dipolar magnetic field is present, and the accretion flow is characterized by a disk truncation radius radius (\ri) and the width of the flow (\rw) at the disk plane. A steady flow prescription occurs for a given {\mdot} to determine the density at a given point. The temperature at each point scales with density with a constant heating rate in the flow; the maximum temperature in the flow (\tmax) describes each model. We use the extended Sobolev approximation and calculate the mean intensity and the level population of a 16-level hydrogen atom and the ray-by-ray method for a given viewing inclination ($i$).

We created a grid of 35,200 models varying \mdot, \ri, \rw, \tmax, and $i$ using the ranges of parameters appropriate for CTTS \citep{muzerolle01}. We convolved the model profiles with a Gaussian instrumental profile of CHIRON's resolution and added a chromospheric + photospheric contribution to the model using an observation of the WTTS TWA~25 (spectral type = M0, \citealt{manara17}) convolved to the same resolution. We then fitted each observed profile inside $\pm400~\kms$ from the line center, and the best fits are determined by calculating the $\chi^2$ for each combination of the model and observed profile. We selected the top 100 models that can simultaneously fit both lines and calculated the weighted mean and standard deviation of \mdot, \ri, \rw, \tmax, and $i$, where we used the likelihood $L=e^{-(\chi^2_{H\alpha} + \chi^2_{H\beta})}$ as the weight. In Figure~\ref{fig:accflow}, we show the best-fitting models to the {\halpha} and {\hbeta} profiles, with the best-fit parameters in Table~\ref{tab:accflow}. The mass accretion rate is consistent with that from accretion shock modeling.

In general, the main features in the line that we aim to fit are the wings, since their emission comes largely from the magnetosphere. Our model fits the wings of the {\halpha} line very well. The small discrepancy at the line center is likely due to other contributions that we cannot fully account for, such as the stellar chromosphere and line emission from the B component.

\begin{deluxetable}{ccc}
\tablecaption{Results of the Extended Accretion Flow Model \label{tab:accflow}}
\tablehead{
\colhead{Parameter} &
\colhead{Value} &
\colhead{Unit}
}
\startdata
{\mdot} & $2.56\pm1.31$ & $10^{-8}$ \msunyr \\
{\ri}   & $2.25\pm0.16$ & $R_\star$ \\
{\rw}   & $0.20\pm0.00$ & $R_\star$ \\
{\tmax} & $7230\pm30$   & K \\
$i$     & $37.0\pm 8.3$ & deg\\
\enddata
\tablecomments{$R_\star = 1.80\,R_{\odot}$. See Table~\ref{tab:starparam}.}
\end{deluxetable}

For the {\hbeta} line, we fit the blue wing reasonably well. However, the extra redshifted emission on the red side is not reproducible by our model. \citet{campbell-white21} suggest that this emission, as well as the rapidly variable redshifted components observed in many lines, could come from rotating infalling material or an inner disk feature, similar to what has been observed in other stars like EX Lupi \citep{sicilia15}. Exploring this possibility would need further time-resolved observations and can be investigated in future work.

The mass accretion rates found by line modeling are in reasonable agreement with that found by accretion shock modeling (Section~\ref{sec:accretionshock}). Furthermore, we calculated the energy flux $\curf$ and the filling factor $f$ of the accretion flow, assuming the best-fitting geometry, and found that $\curf=5.5\times10^{11}~\escm$ and $f=0.047$, respectively, which are in fairly good agreement with those in Table~\ref{tab:outmodelparam}.

The inclination of $\sim$40$^{\circ}$ found from line modeling suggests that the inclination of the magnetic axis may be larger than $i\sim$14$^\circ$ for the stellar and presumably disk, rotation axis. Such misalignments between stellar and magnetic axes are not uncommon \citep[e.g.,][]{McGinnis20,Johnstone14} and would result in Balmer line profiles that vary with rotation phase.

\subsection{Accretion Rate Variability} \label{sec:mdot_variability}
The accretion rate for a given CTTS is expected to vary by a factor of about 0.5 dex \citep{venuti14}. The analysis of the \halpha\ profiles presented in Section \ref{sec:Halpha_spectra} suggests that {\it HST} observed CVSO~109A during a modestly enhanced accretion event. The accretion rates of $\sim 3\times10^{-8}$ \msunyr\ reported in Sections \ref{sec:NUVtoNIR}-\ref{sec:extendedflow} are consistent with previous measurements. \cite{manara21} used a slab model to derive an accretion rate of $3.24\times10^{-8}$ \msunyr\ (contemporaneous to {\it HST}), and \cite{ingleby14} used a five-column shock model to find an accretion rate of $3.0\times10^{-8}$ \msunyr. Both groups analyzed the unresolved CVSO~109 system, but, as discussed in Section \ref{sec:NUVtoNIR}, we expect that accretion-related excess emission comes almost entirely from the CVSO~109A component.

\citet{mauco18} measured an accretion rate of $0.67\times10^{-8}$ \msunyr\ from CVSO 109's \halpha\ line luminosity (estimated as its equivalent width times the continuum flux) using the \halpha-\mdot\ relation found for CTTSs in Taurus by \citet{ingleby13}: log(\mdot)=1.1($\pm0.3$)log(L$_{\rm H\alpha})-5.5(\pm0.8$). To test whether the discrepancy in measured \mdot s results from the different measurement techniques or from real variability, we follow the procedure of \citet{mauco18} in using the contemporaneous (labeled day 0 in Section \ref{sec:Halpha_spectra}) CHIRON EW(H$\alpha$) with the above \halpha-\mdot\ relation to estimate \mdot. Depending on the exact procedure used for fitting the continuum and handling the line wings, we attain EW(\halpha) in the range between -30 and -40 \AA. Using the same distance to CVSO 109 as \citet{mauco18} (440~pc) and taking into account the spread in the \halpha-\mdot\ relation, this range of EW(\halpha) gives \mdot\ between 0.41$\times10^{-8}$ \msunyr\ and 0.67$\times10^{-8}$ \msunyr, fully consistent with the value obtained in \citet{mauco18}. Thus, the discrepancy can be explained by the difference in measurement techniques rather than true variability. 

The $\dot{M}$--age relationship presented in Equation 12 of \citet{hartmann16} gives an accretion rate on the order of a few times $10^{-9}$ \msunyr\ for a 5 Myr TTS. The accretion rates reported for CVSO 109 from modeling its UV continuum excess emission, which are an order of magnitude higher than the empirical estimate, show that CVSO~109A is a strong accretor even with its host region's intermediate age of 5~Myr. High accretion rates at ages of 3--10 Myr are also found elsewhere \citep{ingleby14,rugel18,manara20}, despite the expectation that TTS accretion rates decrease with age
 \citep[e.g.,][]{hartmann98, briceno19}.

\subsection{Absorption by a Wind from the Inner Disk}
\label{sec:InnerDiskWind}

 Wind absorption in CVSO~109A is detected in abundant, low-ionization species with strong transitions that have low-energy lower levels. We first discuss the presence and absence of various lines and then interpret the line profiles.  The wind absorption described in this section occurs close to the star and is analyzed independently of the [\ion{O}{1}] emission (see Sections 6.5.3 and 6.5.4), which traces gas from a distinct wind that is launched over larger radii.  In this section, we compare the wind profile to that of RU Lup, which has an accretion rate of $\sim 7\times10^{-8}$ \citep{alcala17}, a few times higher than CVSO 109A, and face-on disk with an inclination of 18.8$^\circ$ \citep{huang18}.

\begin{figure}
\epsscale{1.15}
\plotone{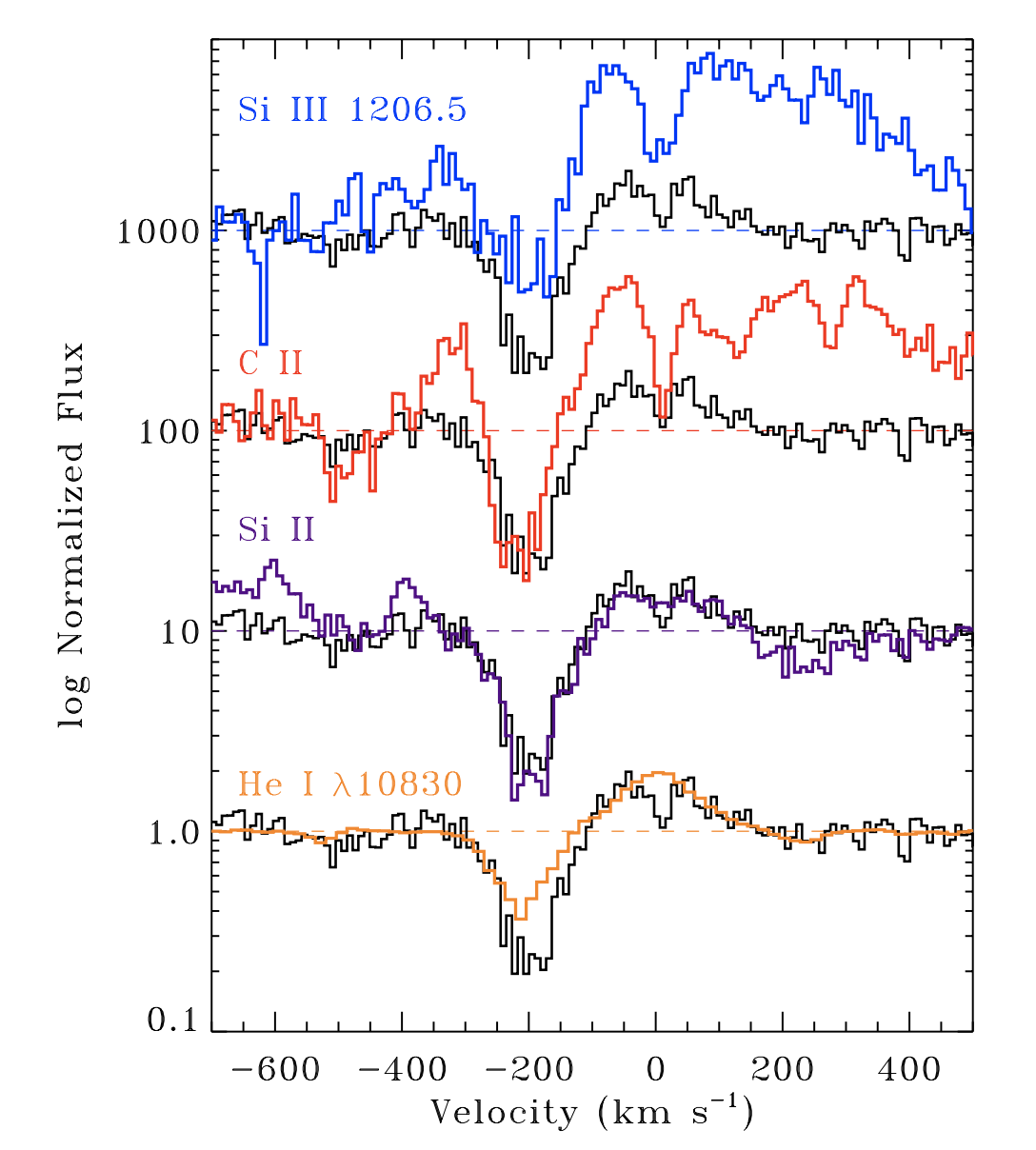}
\caption{Wind profiles from lines of \ion{Si}{2} (coadded), \ion{C}{2} (coadded), \ion{Si}{3} $\lambda1206.5$, and \ion{He}{1} $\lambda10830$ (colors), normalized to the surrounding continuum. Each profile includes a coadded profile (thin black spectrum) that combines the \ion{Si}{2} and \ion{C}{2} lines to help guide the eye. Narrow absorption features near $0~\kms$ trace either interstellar or circumstellar material.  All coadded profiles are combined in flux space.}
\label{fig:windprofs}
\end{figure}

\subsubsection{Wind Absorption Lines}

Blueshifted subcontinuum absorption is detected in \ion{O}{1}, \ion{Al}{2}, \ion{Si}{2}, \ion{C}{2}, and \ion{Si}{3} (see selected lines in Figure~\ref{fig:windprofs}). These wind absorption lines are typical of those detected in FUV spectra of CTTS \citep[e.g.][]{herczeg05,xu21}. Wind absorption is also detected in \ion{He}{1} $\lambda10830$ in our X-shooter spectrum.  

Some lines that are detected in spectra with strong winds, such as RU~Lup \citep{herczeg05,xu21}, are not detected in CVSO~109A. The excited \ion{N}{1} $\lambda\lambda1492,1494$ doublet has a very marginal subcontinuum signal that is not a clear detection. FUV \ion{Fe}{2} lines with excited lower levels are also not clearly detected. The low-resolution NUV spectrum shows strong \ion{Fe}{2} emission in features that for other stars are seen in absorption and can be identified even at low resolution.

The non-detection of wind absorption in \ion{Fe}{2} and possibly in \ion{N}{1} may be a consequence of a lower mass-loss rate than stars that show \ion{Fe}{2} in absorption, differences in the line of sight, or perhaps abundance effects.  The detected wind absorption lines are all ionized, likely by the strong radiation field from the star. The \ion{N}{1} detections might only occur in winds that have sufficient mass-loss rates to provide some shielding to ionizing radiation. The hotter lines, \ion{Si}{4}, \ion{C}{4}, and \ion{N}{5}, show no evidence of any wind absorption against the continuum or against H$_2$ emission lines.


\begin{figure}
\epsscale{1.15}
\plotone{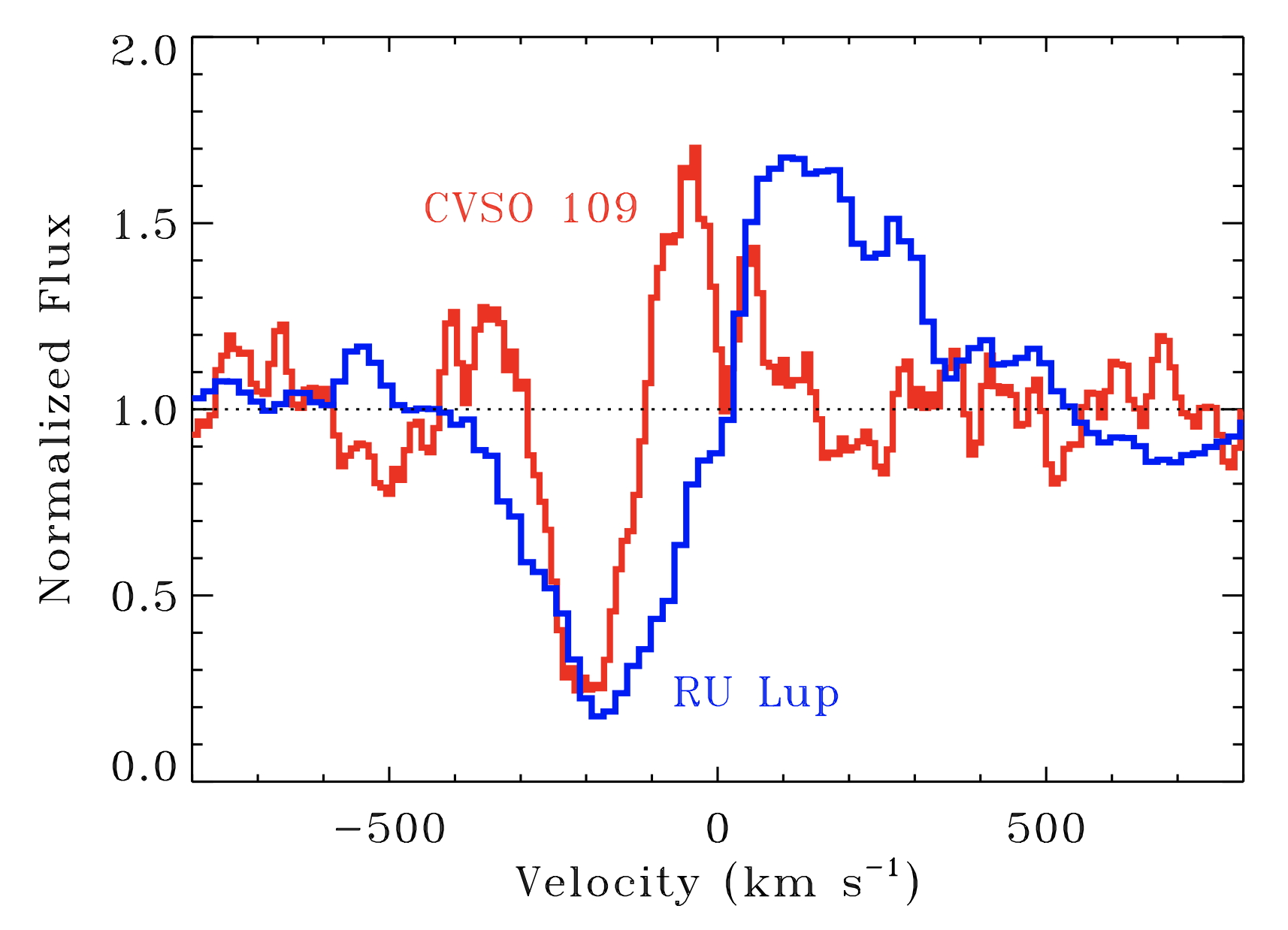}
\caption{The wind profile from CVSO~109A, obtained from coadded \ion{C}{2} and \ion{Si}{2} lines (\ion{C}{2} lines at 1334.5 and 1335.7 \AA\ and \ion{Si}{2} lines at 1260.422, 1264.738, 1526.707, 1533.431, 1194.5, 1197.394, 1193.290, and 1190.416 \AA), compared with the wind from RU~Lup, as seen with COS \citep[data obtained from][]{france12}. The wind from RU~Lup starts at lower velocity and extends to higher velocity, while the wind absorption from CVSO~109A is narrower, has a faster minimum velocity, and is centered at $-220~\kms$.}
\label{fig:windprofcomp}
\end{figure}

\subsubsection{Wind Absorption Profiles}

In the framework for wind absorption lines, as developed for \ion{He}{1} $\lambda10830$ by \citet{edwards06}, fast absorption occurs from a wind launched either from the star alone or the star and disk interaction region, while slow absorption traces a disk wind. Figure~\ref{fig:windprofcomp} shows a coadded wind profile for strong \ion{C}{2} and \ion{Si}{2} lines compared with the same coadded line for RU~Lup. The CVSO~109A wind extends from about $-100$ to $300~\kms$, leading to a classification as a fast wind. The minimum velocity of $\sim100~\kms$\ is sensitive to the underlying emission profile and is difficult to measure accurately. Most fast winds are detected from disks that are viewed close to face-on. The fast wind, with a maximum velocity of about $-300~\kms$\, is consistent with the maximum velocities measured in \ion{C}{2} lines from star-disk systems that are viewed at low inclination
 \citep{xu21}. 

The absorption line shapes are somewhat unusual at low velocity.  Most fast winds have minimum detected velocities\footnote{the minimum velocity that is clearly below the continuum or expected emission line.} between
$0$ to $-50$~$\kms$ \citep{edwards06,xu21}.  
For CVSO 109A, the minimum velocity with clear wind absorption is about 
 about $-100$~$\kms$. 
 We speculate that for most other TTS with fast winds, the line of sight passes through the acceleration region, leading to absorption across a wide range of velocities. For CVSO~109A, the line of sight may not intercept the base of the wind, where the gas is traveling slowly. This interpretation requires that the wind is magnetized and accelerates with distance from the launch region \citep[see review by][]{frank14}.

\subsubsection{Estimating the Mass-Loss Rate from the Inner Wind}

Based on the blueshifted absorption component in the {\henir} line, we can make an estimate of the mass-loss rate from the CVSO~109A system, following \citet{calvet97} and \citet{thanathibodee20}. For lines in Figure~\ref{fig:windprofs}, the blueshifted absorption components are at high velocities. Therefore, we can adopt the Sobolev approximation to relate the optical depth $\tau(v)$ at velocity $v$ and local properties in the wind, giving 
\begin{equation}
    \tau(v) = \frac{\pi e^2}{m_ec}\frac{fc}{\nu_0}\frac{n_l(v)}{dv/dz},
\end{equation}
where $f$ is the oscillator strength of the line with line-center frequency $\nu_0$, $n_l$ is the number density of the lower level of the line, and $dv/dz$ is the velocity gradient. Knowing $\tau$ from the observation, $n_l$ can be determined, and the mass-loss rate is calculated by
\begin{equation}
    \dot{M}_w \sim \Delta A v \mu \eta n_l(v),
\end{equation}
where $\Delta A$ is the cross-sectional area of the wind, $\mu$ is the mean molecular weight, and $\eta \equiv n_H/n_l$ is the ratio between the number density of hydrogen and that of the lower level of the line. Following \citet{calvet97}, we approximate $dv/dz\sim v/R_{\star}$ and $\Delta A\sim\pi(2R_{\star})^2$.

From Figure~\ref{fig:windprofs}, we have $\tau=-\ln(0.36)=-1.02$ at $v=-220\,\kms$, where the blueshifted absorption is deepest. Therefore, we have $\dot{M}_w\sim3.2\times10^{-10}(\eta/10^7)$\,\msunyr. The parameter $\eta$ is uncertain and depends on the wind density and temperature and the radiation field. \citet{thanathibodee20} estimated $\eta\sim10^7$ for the star PDS~70 from irradiating a slab of gas with an X-ray source emitting as a blackbody. If the measured $L_X$ is split equally between the two components of CVSO~109, then we could assign $L_X\sim1.5\times 10^{30}\,{\rm erg~s^{-1}}$ to each. With an assumed effective temperature $T_x\sim5\times10^6$\,K and $\log(n_H/{\rm cm}^3)\sim8$--10, we obtain $9\times10^6\lesssim\eta\lesssim2\times10^8$. Therefore, the mass-loss rate estimated from the {\henir} line is between $3\times10^{-10}$ and $6\times10^{-9}$\,\msunyr.

\subsection{FUV Radiation Environment}
\label{sec:FUV_radiation_environment}

The FUV radiation field of accreting young stars comprises a continuum and bright line emission generated near the accretion shock. Previous studies have shown that the continuum and the accretion-dominated Ly$\alpha$ emission make up the vast majority of the disk-illuminating FUV radiation field~\citep{herczeg04, schindhelm12, france14}. 
The typical FUV spectrum of a CTTS also includes emission from H$_{2}$ and CO lines and continuum~\citep{france11b}; however, these molecular emissions are generated in the surrounding disk and/or winds around the accreting protostar and are typically not included in the FUV radiation budget emitted from the stellar region. We also measure the size of the H$_{2}$ emitting region and analyze line profiles for UV-H$_{2}$ and optical wavelength [\ion{O}{1}]  profiles, both of which are expected to arise near the disk surface, possibly part of a slow disk wind~\citep{france12,gangi20}.  

We measure the contribution of these different stellar and accretion components using the spectral decomposition described in~\citet{france14}. First, we measure the observed fluxes from hot gas lines in the spectra (e.g., \ion{C}{4}). Second, we spectrally isolate the FUV continuum from the many narrow fluorescence lines and molecular continuum \citep[noting that the molecular continuum in CVSO~109A has been shown to be weak or absent;][]{france17}. Third, we use the observed H$_2$ fluorescence lines to reconstruct the Ly$\alpha$ radiation field impinging on the disk surface (\citealt{herczeg04,schindhelm12, arulanantham18}, following \citealt{wood02}). We also measure the size of the H$_2$ emitting region and analyze the UV-H$_{2}$ and [O~I] emission profiles.

In Section 6.5.1 and 6.5.2, we describe the measurement and characteristics of the FUV continuum and Ly$\alpha$ emission from CVSO 109A, and then present the relative contributions to the FUV radiation field in Section 6.5.3.  Section 6.5.4 describes the disk emission from fluorescent H$_{2}$.   We compare the H$_{2}$ fluorescence spectrum  to the [\ion{O}{1}] disk wind in Section 6.5.5.

\subsubsection{FUV Continuum of CVSO~109}

The FUV spectrum of CVSO 109A is qualitatively similar to other CTTS analyzed in the literature, comprising strong neutral and intermediate ionization atomic lines (e.g., \ion{H}{1} Ly$\alpha$~--~see Section 6.5.2, \ion {C}{2}~--~see Section 6.4, \ion{C}{4}~--~see Section 6.2), fluorescent H$_{2}$ emission (see Sections 6.5.2 and 6.5.4), where vibrationally-excited molecules are ``pumped'' by the strong Ly$\alpha$ radiation field incident on the inner disk, and continuum emission extending from $\sim$~1100~--~1750~\AA. Unlike some accreting young stars, the spectrum of CVSO 109A does not include emission from fluorescent CO~\citep{france11b, arulanantham21} or the highly-excited H$_{2}$ continuum emission that is possibly produced during the dissociation of water molecules~\citep{france17}.  

\begin{figure}    
\epsscale{1.15}
\plotone{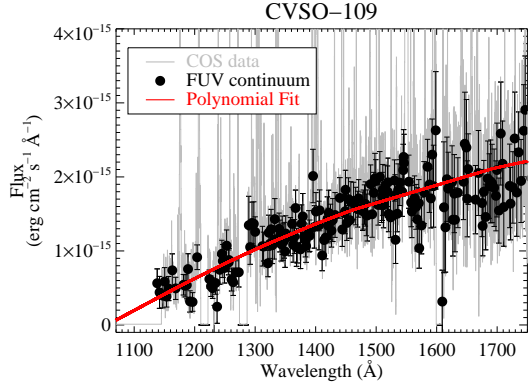}
\vspace{-0.3in}
\caption{The FUV continuum spectrum of CVSO~109 A. The full COS spectrum is shown in gray while the black filled circles are the average fluxes in 0.75~\AA\ bins selected to exclude bright stellar and disk emission lines. A second-order polynomial fit is shown in red.
}
\label{fig:sfuvcont}
\end{figure}

To measure the FUV continuum, we require data with moderate to high spectral resolution so as to separate the lines from the continuum, while also having sufficient sensitivity to measure continuum fluxes in the spectrally narrow ($\sim0.75$~\AA) bins characteristic of the interline regions of CTTS spectra. We manually define 205 continuum points with $\Delta\lambda=0.75$~\AA\ bins ranging from 1138 to 1791~\AA; the continuum spectral data points are the average flux in each bin, and the uncertainty is defined as the standard deviation of the flux in each continuum bin. The full FUV continuum spectrum of CVSO~109 A (Figure~\ref{fig:sfuvcont}) is fit with a second-order polynomial extending down to  912~\AA\ to estimate the flux levels in the H$_2$-dissociating and CO-dissociating radiation field below 1110~\AA. We measure an integrated 912--1650~\AA\ continuum flux of $6.7\times10^{-13}~\escm$. Assuming the 400~pc distance for CVSO~109, this gives an FUV continuum luminosity of $1.3\times10^{31}$ erg~s$^{-1}$.   

\subsubsection{Ly$\alpha$ Emission}

We used the modeling framework developed by \citet{schindhelm12} and electronic transition rates from \citet{abgrall93} to reconstruct the Ly$\alpha$ profile of CVSO~109 A from the integrated fluxes of fluorescent H$_2$ emission lines. The fluorescent H$_2$ molecules producing the emission can reach excited electronic states only through radiative pumping, via photons at energies that span the full width of the Ly$\alpha$ profile. These upper electrovibrational levels are pumped out of ground electronic levels ($X$$^{1}\Sigma^{+}_{g}$, $v''$, $J''$) with radiative transitions that overlap with the strong Ly$\alpha$ emission line (see, e.g., Wood et al. 2002 and Herczeg et al. 2002 for a detailed description).   Since observed Ly$\alpha$ fluxes near the line center are attenuated by interstellar and circumstellar absorption and geocoronal emission, molecular fluorescence lines are often the only accessible tracers of the full Ly$\alpha$ radiation field reaching the disk surface.

\begin{figure}    
\epsscale{1.15}
\plotone{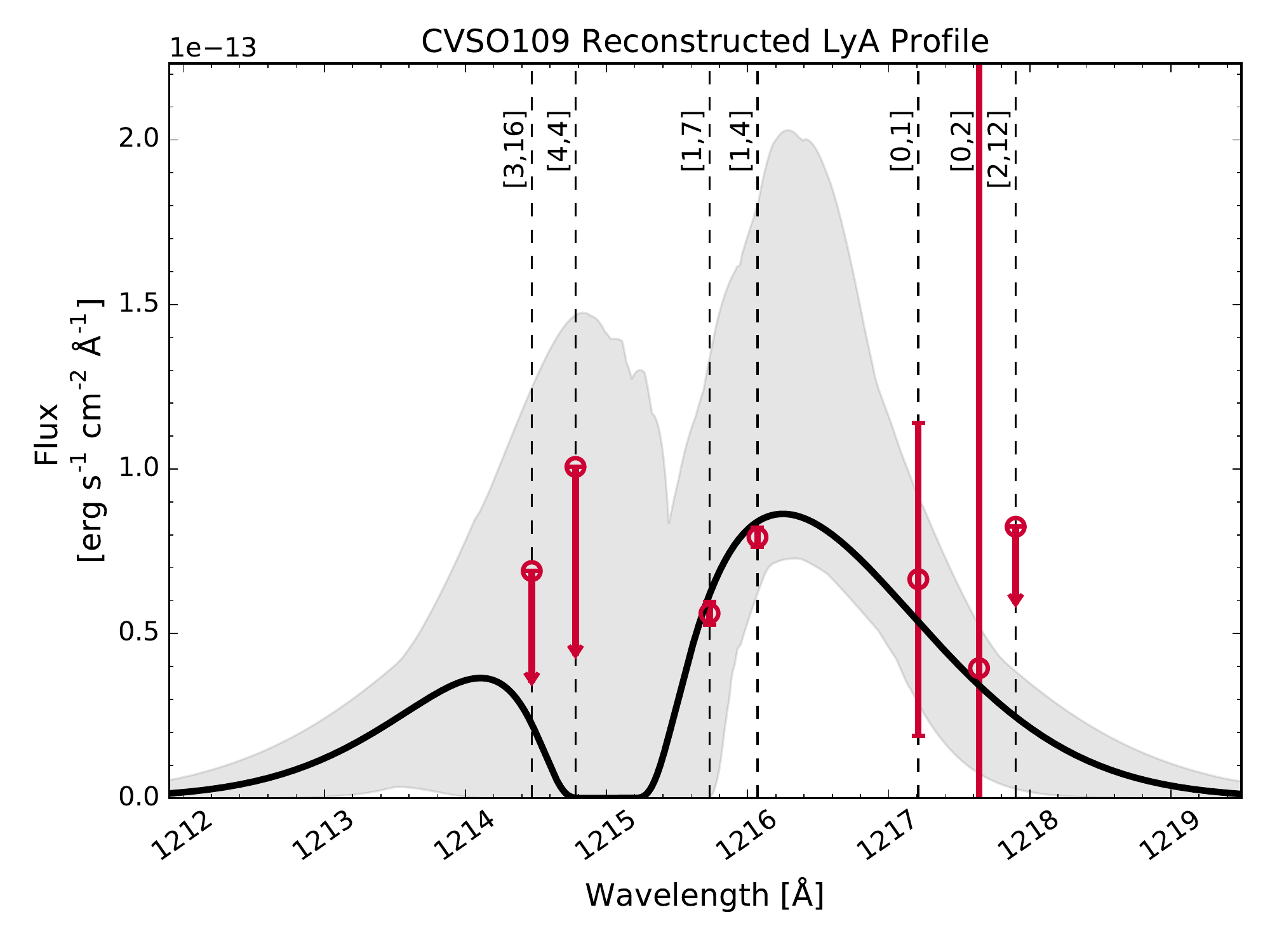}
\caption{The best-fit reconstructed Ly$\alpha$ profile of CVSO~109 A (black, solid curve), with $\pm1\sigma$ contours drawn based on the uncertainties in the model parameters (gray, shaded regions). The model was fit to “data points” constructed from fluorescent H$_2$ emission line fluxes (red circles), which allowed us to trace the Ly$\alpha$ reaching the disk surface at wavelengths where the observed profile is attenuated by geocoronal emission and interstellar absorption. Since CVSO~109 A shows detectable H$_2$ emission lines from only four upper levels redward of the Ly$\alpha$ line center, the blue side of the reconstructed profile is less well constrained, as seen most clearly in the wide spread of possible \ion{H}{1} outflow velocities.}
\label{fig:reconstLyA}
\end{figure} 

\begin{figure}    
\epsscale{1.15}
\plotone{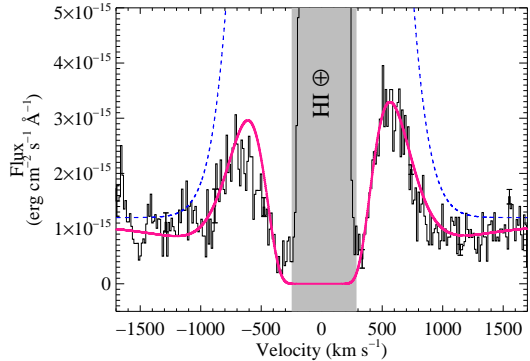}
\vspace{-0.3in}
\caption{The accretion-dominated Ly$\alpha$ and interstellar \ion{H}{1} attenuation profile of CVSO~109 A. The COS data are the gray histogram, the intrinsic Ly$\alpha$ emission line is the blue dashed line, and the pink  solid line is the best fit model of the stellar emission line and the interstellar absorption profile. The best-fit interstellar hydrogen column density is $\log_{10}(N($\ion{H}{1}$))=20.32\pm0.06$.}
\label{fig:slya1}
\end{figure} 

In Figure~\ref{fig:reconstLyA}, the H$_2$ fluxes were first used to construct new “data points” to be fit with the Ly$\alpha$ reconstruction model. $x$ values of the new data points are the Ly$\alpha$ wavelengths required to pump the molecules to upper levels in excited electronic states. Corresponding $y$ values are the sums of integrated fluxes from all observed H$_2$ features with the same upper level, with each individual emission line flux weighted by a branching ratio that describes the probability of the transition relative to all other pathways out of the upper level. Although this approach is typically used to build a total of 12 data points, H$_2$ emission lines from CVSO 109 were clearly detected from only 4 upper levels of the $B$$^{1}\Sigma^{+}_{u}$ excited electronic level: $[v’,J’]=[1,4]$ ($x=1216.07$~\AA); [1,7] ($x=1215.726$~\AA); [0,1] ($x=1217.205$~\AA); and [0,2] ($x=1217.643$~\AA).   Since all detected emission lines are pumped by Ly$\alpha$ photons redward of line center, this leaves the blue side of the reconstructed profile unconstrained (although we note that the blue side of the line is directly observed outside of the interstellar neutral hydrogen line core, see Figure~\ref{fig:slya1}). We list the central wavelengths of the detected features in Table \ref{tab:LyAH2}.

\begin{deluxetable}{cccc}
\tablecaption{H$_2$ Emission Lines Used for Ly$\alpha$ Reconstruction \label{tab:LyAH2}}
\tablehead{
\colhead{[$v'$, $J'$]} &
\colhead{$\lambda_{\rm{pump}}$} &
\colhead{$\lambda_{\rm{H}_2}$} & \colhead{$B_{mn}$}
}
\startdata
[1,7] & 1215.726 \AA & 1467.08 & 0.08 \\
\nodata & \nodata & 1500.45 & 0.101 \\
\nodata & \nodata & 1524.65 & 0.111 \\
\nodata & \nodata & 1556.87 & 0.074 \\
\hline
[1,4] & 1216.07 \AA & 1431.01 & 0.058 \\
\nodata & \nodata & 1446.12 & 0.083 \\
\nodata & \nodata & 1489.57 & 0.094 \\
\nodata & \nodata & 1504.76 & 0.115 \\
\hline
[0,1] & 1217.205 \AA & 1398.95 & 0.141 \\
\nodata & \nodata & 1460.17 & 0.083 \\
\hline
[0,2] & 1217.643 \AA & 1402.65 & 0.126 \\
\nodata & \nodata & 1463.83 & 0.074 \\
\nodata & \nodata & 1525.15 & 0.029 \\
\enddata
\tablecomments{[$v'$, $J'$] describes the upper level of the transitions excited by a given Ly$\alpha$ wavelength $\left( \lambda_{\rm{pump}} \right)$. $\lambda_{\rm{H}_2}$ values are the wavelengths of detected H$_2$ emission lines from upper level [$v'$, $J'$], and $B_{mn}$ are the branching ratios describing the transition probabilities $\left( B_{mn} = A_{v', J' \longrightarrow v'', J''} / A_{v', J'} \right)$}.
\end{deluxetable}

The four data points produced from the H$_2$ emission line fluxes were then fit with a model consisting of an intrinsic Gaussian emission line centered at the stellar velocity and a superimposed blueshifted Voigt absorption profile from outflowing \ion{H}{1}. Variable parameters were the Gaussian amplitude $\left(I_0 \right)$ and FWHM, characteristic outflow velocity $\left(v_{\rm out}\right)$, and \ion{H}{1} column density of the blueshifted outflow $\left(N\left(\rm{H\,I}\right)_{\rm out}\right)$. To reproduce the $y$ values, we assumed that Ly$\alpha$ photons at the disk surface were absorbed by a thermal population of H$_2$ and varied the temperature $\left(T_{\rm{H}_2} \right)$ and column density $\left( N \left(\rm{H}_2 \right) \right)$ of this distribution. The best-fit parameters are $I_0=(2\pm1)\times10^{-13}~\escm$~\AA$^{-1}$, $\rm{FWHM}=600^{+300}_{-200}~\kms$, $T_{\rm{H}_2}=4200^{+700}_{-1500}$~K, $N \left(\rm{H}_2\right)=21.6^{+0.3}_{-1.3}$~dex, $v_{\rm out}=-140\pm100~\kms$, and $N\left(\rm{H\,I}\right)_{\rm out}=18.9^{+0.9}_{-1.5}$~dex.  Although this outflow velocity appears qualitatively consistent with the central velocity of the observed wind profile (Figure \ref{fig:windprofcomp}), any comparison is limited by uncertain wind geometry and the differences in the lines-of-sight. The total integrated model Ly$\alpha$ flux is $F(\rm{Ly}\alpha)_{recon}=\left(2.7\pm0.8\right)\times10^{-13}~\escm$ over the reconstructed, outflow-absorbed, profile, which is consistent with the total flux derived from the ISM model (see below). The FWHM of the reconstructed profile is consistent with the average from the sample modeled in \citet{schindhelm12} $\left(\rm{FWHM}\sim700~\kms\right)$ as well as the observed Ly$\alpha$ line wings described in the next paragraph. 

The high line flux of Ly$\alpha$ enables its use as a background source against which neutral hydrogen in the foreground ISM can be measured (see, e.g., McJunkin et al. 2014 and references therein.
We therefore used the observed suppression of the red wing of the Ly$\alpha$ emission line to estimate the interstellar hydrogen column toward CVSO~109 A.\footnote{The blue wing of the observed Ly$\alpha$ profile is affected by high-velocity neutral outflow from the stellar region~\citep{schindhelm12, mcjunkin14,arulanantham21}, and therefore we do not consider it in the ISM fitting.} Given the large uncertainty on the H$_{2}$-derived Ly$\alpha$ line shape, we created a fit to the intrinsic Ly$\alpha$ emission spectrum and interstellar \ion{H}{1} attenuation by assuming a single Gaussian stellar and accretion Ly$\alpha$ emission line and single ISM absorber. The best-fit parameters (Figure~\ref{fig:slya1}) found a Gaussian amplitude of $(7.5\pm1.0)\times10^{-14}~\escm$~\AA$^{-1}$, a ${\rm FWHM}=750\pm100~\kms$, and an emission line RV of $-15\pm10~\kms$. Given that the single Gaussian treatment is an oversimplification of the intrinsic Ly$\alpha$ line profile, the individual emission line parameters and associated uncertainties should be treated with caution. The interstellar hydrogen column density is $\log_{10}(N($\ion{H}{1}$))=20.32\pm0.06$. Coupled with interstellar gas-to-dust ratios~\citep{diplas94} and assuming a Milky Way extinction curve described by $R_V=3.1$~\citep{ccm89}, this atomic hydrogen column corresponds to a visual reddening of $A_V=0.13$. This \ion{H}{1}-based visual extinction level is in agreement with the reddening estimates of $A_V\sim0.06^{+0.24}_{-0.24}$ derived in Section~\ref{sec:stellarproperties}.
 
\subsubsection{FUV Radiation Field Contributions}

Based on an FUV spectral analysis of 14 CTTS with high-resolution spectra from {\it HST}, \citet{france14} present (1) FUV radiation field component fluxes, normalized to a distance of 1~au from the central star, and (2) fractional contributions of \ion{C}{4}, the FUV continuum, and Ly$\alpha$ to the FUV radiation field. These two metrics present normalized flux measures against which we can compare the FUV radiation properties of CVSO~109 A. For CVSO~109 A, the radiation field component fluxes at 1~au are: \ion{C}{4}, $4.6\times10^{2}~\escm$; FUV continuum, $4.6\times10^{3}~\escm$; and Ly$\alpha$, $2.4\times10^{-13}~\escm$. The integrated FUV flux from CVSO~109 A at 1~au from the star is approximately $6.7\times10^{3}~\escm$, which is very typical of other accreting protostars with a range of central star masses \citep{france14}.

The average CTTS flux contribution fractions over the 912--1650~\AA\ band are F(continuum)/F(FUV) = 8.4\%, F(Ly$\alpha$)/F(FUV) = 88.1\%, and F(\ion{C}{4})/F(FUV) = 2.1\% \citep{france14}\footnote{ Contributions from weaker atomic emission lines, e.g., \ion{N}{5} and \ion{He}{2}, make up the remaining 1.4\% of the FUV radiation field budget.}. For CVSO~109 A, we find an unusually small Ly$\alpha$ contribution (or, alternately, unusually large continuum and hot gas contributions) to the total FUV flux:  F(continuum)/F(FUV) = 68.5\%, F(Ly$\alpha$)/F(FUV) = 24.6\%, and F(\ion{C}{4})/F(FUV) = 6.8\%. We suggest that two factors may be at play in the low Ly$\alpha$ and high continuum fractions seen from CVSO~109 A. First, it may be that CVSO~109 A generates less intrinsic Ly$\alpha$ flux than other accreting TTS. Second, the lack of CO fluorescence and ``1600~\AA\ bump’’ emission in CVSO~109 A, both species which are pumped by Ly$\alpha$ photons, indicates that Ly$\alpha$ photons do not penetrate deeply into the disk environment; the neutral gas and dust opacity associated with a primordial gas-rich disk may be similarly attenuating the observed Ly$\alpha$ emission. 

Low Ly$\alpha$ disk irradiance is unusual for an older CTTS \citep[$\sim5$~Myr,][]{briceno19}. Recent molecular disk surveys have found that more evolved disks display signs of greater Ly$\alpha$ propagation and influence on disk excitation, as seen from H$_{2}$O dissociation tracers and CO fluorescence~\citep{france17,arulanantham21}.

Figure~\ref{fig:reconstLyA} includes the upper limits on H$_2$ emission line fluxes from three undetected progressions, pumped at 1214.47, 1214.78, and 1217.9 \AA\ ($[v',J']=[3,16]$, [4,4], and [2,12]). 
However, only 3/27 targets from \citet{france12} do not show detectable [3,16] or [4,4] features, and only 2/27 systems do not have [2,12] emission lines. To explore whether the nondetections in CVSO~109 A are unusual, we estimate the expected [4,4] flux using the linear relationship between the ratios of red to blue wing observed Ly$\alpha$ emission and red ($[1,4]+[1,7]$) to blue ([4,4]) H$_2$ progression fluxes from \citet{arulanantham21}. The expected [4,4] emission from CVSO~109 A is $\sim5.6\times10^{-15}~\escm$, which sits well below the measured upper limit of $8.2\times10^{-14}~\escm$ and detected [4,4] fluxes from other T Tauri stars (median $F_{\rm{[4,4]}} \sim 7 \times 10^{-14}$ erg s$^{-1}$ cm$^{-2}$; \citealt{france12}) but very close to the Ly$\alpha$ model in Figure~\ref{fig:reconstLyA}. 

\begin{figure}    
\epsscale{1.15}
\plotone{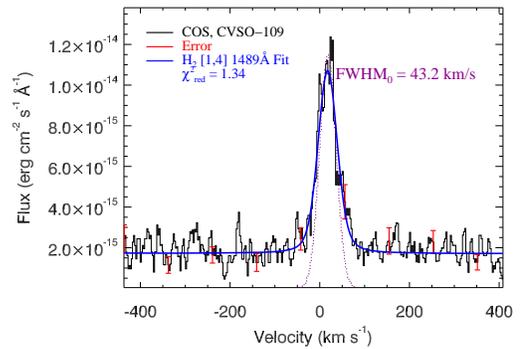}
\vspace{-0.3in}
\caption{The H$_{2}$ (1--7)R(3) emission line pumped by the core of the stellar or  accretion Ly$\alpha$ emission line. The observed line width (FWHM shown in the upper left) indicates the average H$_{2}$ emitting radius between 0.06--0.6~au, depending on the disk inclination (Section~6.5.2).
}
\label{fig:slya2}
\end{figure}

\subsubsection{H$_2$ Emitting Region}

The average fluorescent H$_2$ emitting radius, $\langle R_{H2}\rangle$, can be computed using measurements of the emission line's FWHM, the stellar mass, and disk inclination \citep{france12,arulanantham21}. The rotationally-broadened emission lines can be used to estimate the emitting radius if the gas is bound to the disk or in a slow disk wind (see Section 6.5.5) as that gas retains the kinematic signature along the direction of its orbital motion, e.g., Gangi et al. (2020 and references therein). Following on previous work, we focus on the $B$--$X$ (1--7)R(3) emission line at 1489.57~\AA, from the $[v',J']=[1,4]$ progression pumped near the Ly$\alpha$ line core ($\lambda_{\rm pump}=1216.07$~\AA). Using a Gaussian line-fitting routine that incorporates the extended wings of the COS line spread function (LSF), we measure ${\rm FWHM}_{H2,[1,4]}=43.2\pm2.0~\kms$ (Figure~\ref{fig:slya2}). Lines widths from other H$_{2}$ progressions in CVSO~109 A are similarly in the range $\sim42$--52~$\kms$, typical for H$_{2}$ emission profiles from young stars \citep{france12}. Taking the stellar mass from Table~3 (0.5~{\msun}) and the range of inclination values presented in Section 4.4.3 (14--53\arcdeg), we find an average H$_{2}$ [1,4] emitting radius between 0.06 and 0.6~au. This range of H$_{2}$ emitting radius is typical of average emitting radii for TTS disks and supports the general picture of a gas-rich disk orbiting the primary star.

\subsubsection{UV-H$_2$ and [\ion{O}{1}] Emission Profiles}

Forbidden line emission has long been recognized as a characteristic feature of optical CTTS spectra. Recent surveys of [\ion{O}{1}]  6300~\AA\ emission have drawn tighter connections between these forbidden lines and outflows and disk winds, with multiple-component velocity structures being connected to different emitting and wind-launching radii \citep[e.g.,][and references therein]{simon16}. These winds may be intimately related to the mass accretion process and the long-term dispersal of the protoplanetary disks~\citep{nisini18}. In order to place new constraints on the origin of both forbidden line and molecular disk and wind emission, we compare the [\ion{O}{1}] and H$_{2}$ line profiles of CVSO~109 A in this subsection.

\begin{figure}[t]  
\begin{center}
\epsscale{1.2}
\plotone{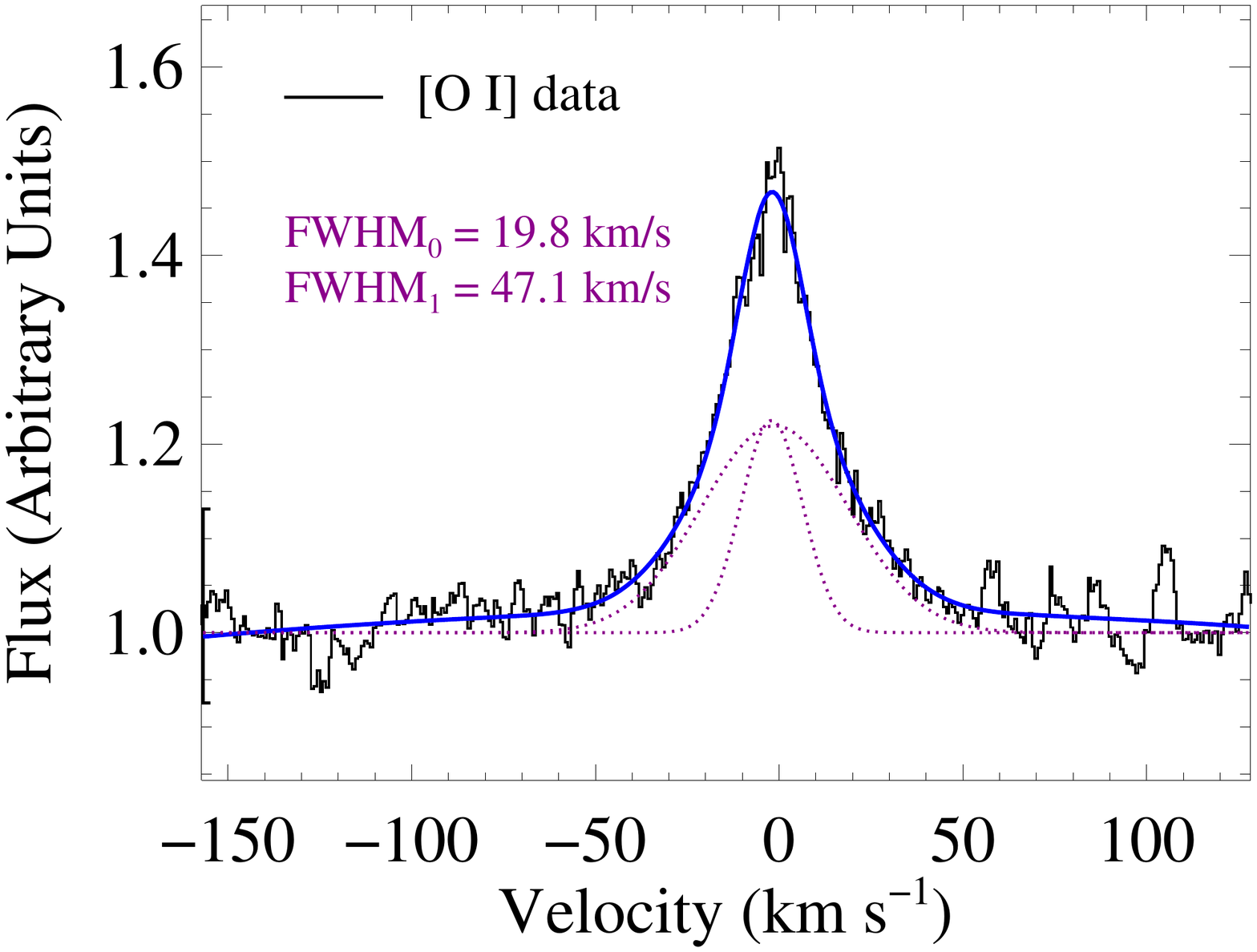}
\vspace{-0.0in}
\plotone{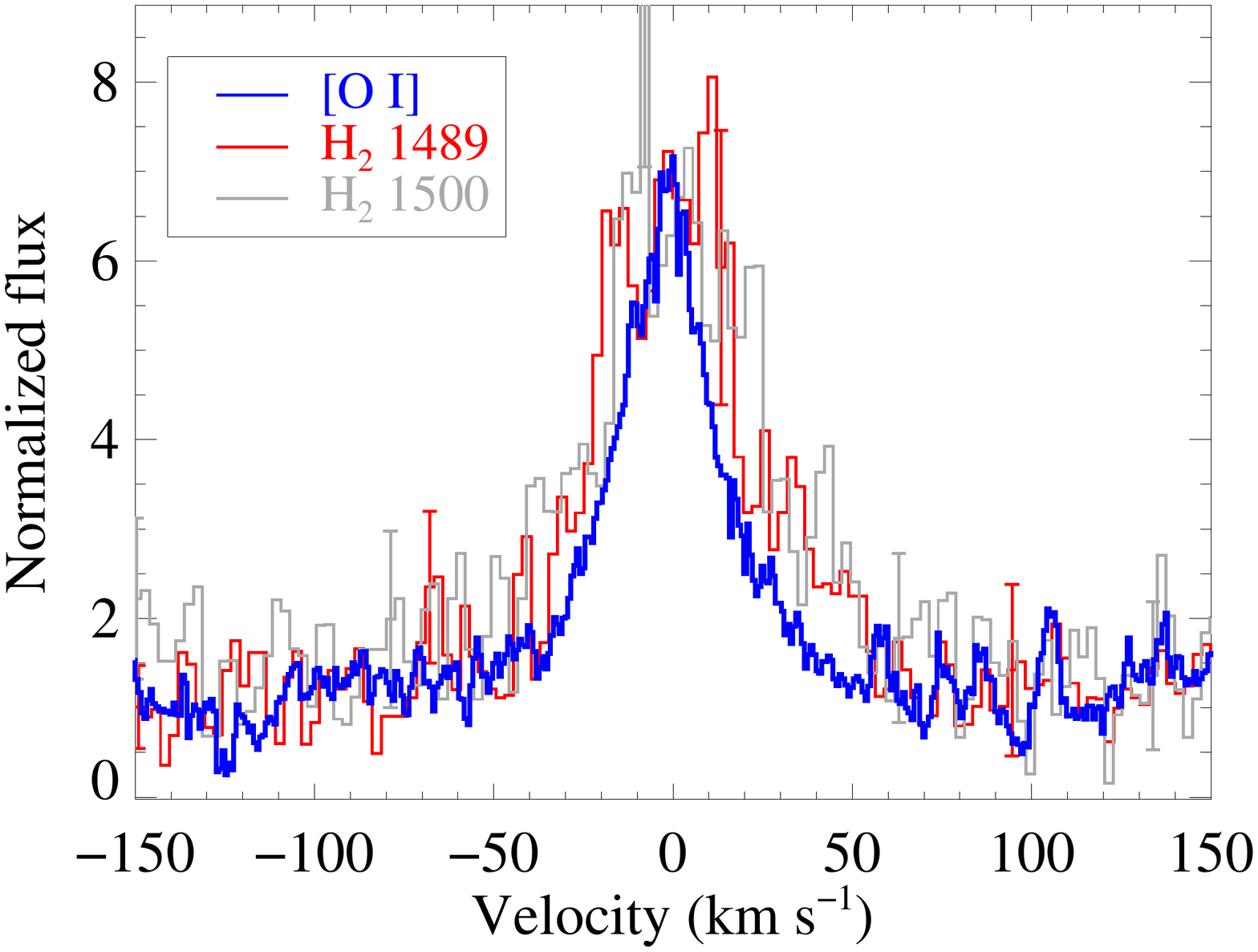}
\vspace{-0.5in}
\caption{ ($top$ $panel$) The two-component fit to the [\ion{O}{1}] 6300~\AA\ emission from CVSO 109A is illustrated.  ($bottom$ $panel$)  Emission profiles from H$_2$ [1,4] 1489~\AA\ (red) and [1,7] 1500~\AA\ (gray) compared with [\ion{O}{1}] 6300~\AA\ (blue). The H$_2$ emission lines have widths of $\sim42$--$52~\kms$ and are consistent with a single component whereas the [\ion{O}{1}] profiles are better fit with a two-component model with FWHM([\ion{O}{1}])$_{\rm narrow}=19.8~\kms$ and FWHM([\ion{O}{1}])$_{\rm broad}=47.1~\kms$.
}
\label{fig:oi_vs_h2}
\end{center}
\end{figure}

Figure~\ref{fig:oi_vs_h2} shows a comparison of the [\ion{O}{1}] $\lambda$6300.304~\AA\ profiles with those of the bright lines of the [1,4] ($\lambda1489.57$~\AA) and [1,7] ($\lambda1500.45$~\AA) progressions of Ly$\alpha$-pumped H$_{2}$. All emission lines have been offset to the rest velocity of CVSO~109 A. As noted above, the width of the [1,4] line is ${\rm FWHM}_{H2,[1,4]}=43.2\pm2.0~\kms$ at an RV of $v_{\rm rad}(H_2[1,4])=+1.5~\kms$. The width of the [1,7] line is ${\rm FWHM}_{H2,[1,7]}=51.6\pm2.8~\kms$ at an RV of $v_{\rm rad}(H_{2}[1,7])=+2.0~\kms$. The [\ion{O}{1}] line is well described by a superposition of two Gaussian emission lines;\footnote{The [\ion{O}{1}] emission could also be fit with a single Gaussian profile centered at $-1.8~\kms$ and with a FWHM of $33~\kms$. This fit leads to larger residuals.} FWHM([\ion{O}{1}])$_{\rm narrow}=19.8~\kms$ at $v_{\rm rad}$([\ion{O}{1}])$_{\rm narrow}=-2.0~\kms$ and FWHM([\ion{O}{1}])$_{\rm broad}=47.1~\kms$ at $v_{\rm rad}$([\ion{O}{1}])$_{\rm broad}=-1.2~\kms$. This [\ion{O}{1}] spectral composition is typical of the ``broad component + narrow component''-type morphology \citep{banzatti19}.  

We find that the RVs of H$_{2}$ and [\ion{O}{1}] are consistent within the $\pm7.5~\kms$ wavelength solution accuracy of {\it HST}--COS. A correlation between the RV of [\ion{O}{1}] and that of H$_2$ was found with the NIR H$_2$ transitions \citep{gangi20}. The H$_2$ line widths are significantly larger than the [\ion{O}{1}] narrow component and similar to the broad component. An important caveat in this comparison relates to the impacts of instrument orientation and spectrograph aperture size. {\it HST}--COS's Primary Science Aperture is a circular opening with a projected angular size of 2.5\arcsec, so contributions from extended H$_2$ emission along the dispersion direction may produce line widths beyond those generated by the kinematic motions of the gas alone. In a few cases, the on-source H$_2$ emission measured with STIS is seen to be blueshifted and in a slow wind \citep[e.g.][]{herczeg06,france12} or located in outflow-cloud interactions \citep{saucedo03,walter03}. The comparison between [\ion{O}{1}] vs.\ FUV H$_2$ line profiles in a larger sample of Orion sources will be presented in Gangi et~al.\ (in preparation).

\begin{figure*}[ht]    
\epsscale{1.1}
\plotone{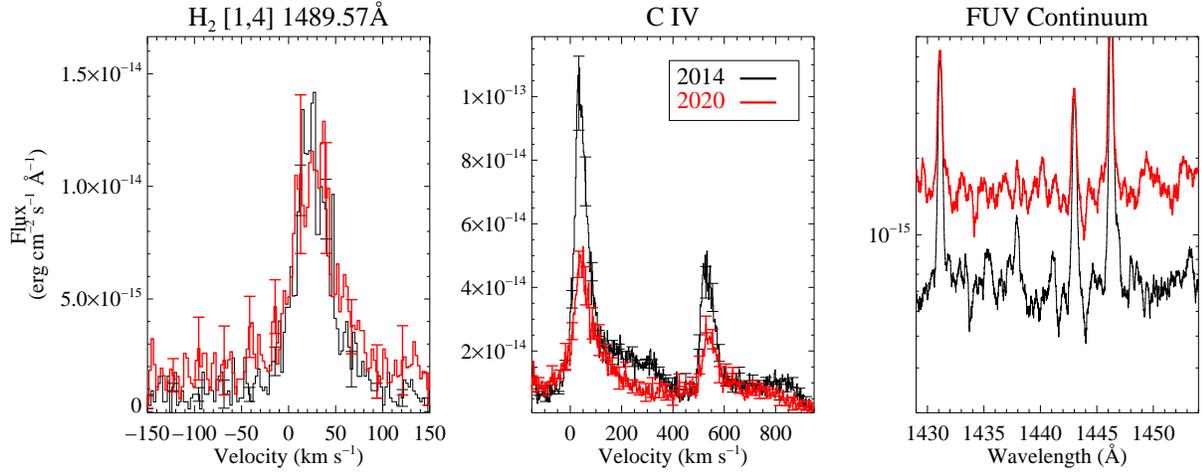}
\vspace{-0.5in}
\caption{A comparison of the 2014 and 2020 disk and accretion profiles of CVSO~109 A. The disk H$_{2}$ profiles are nearly unchanged while the {\civ} and underlying FUV continuum lines show significant temporal variability.
}
\label{fig:fuv_variability}
\end{figure*}

Following the method described in \citet{fang18}, we estimate the wind mass-loss rate traced by [\ion{O}{1}]\,$\lambda$6300, assuming the disk inclination of $\sim40^{\circ}$ derived via line modeling in Section~\ref{sec:extendedflow}. We use as wind velocity ($V_{\rm wind}\sim2~\kms$) projected peak velocities of [\ion{O}{1}]\,$\lambda$6300 and take as wind height ($l_{\rm wind}\sim1$~au) the Keplerian radius from half of the line FWHM, corrected for the instrumental broadening and projected. The wind mass-loss rate ($\dot{M}_{\rm wind}$) is calculated as follows:
\begin{equation}
\dot{M}_{\rm wind}=M\frac{V_{\rm wind}}{l_{\rm wind}}\label{Equ_wind} 
\end{equation}\label{Equ:wind-mass-loss}
\noindent where $M$ is the total mass of gas from the total number of \ion{O}{1} atoms, assuming the oxygen abundance $\alpha({\rm O})=3.2\times10^{-4}$ and the gas temperature $T_{\rm gas}=5000$~K. 
The estimated $\dot{M}_{\rm wind}$ is $1.8\times10^{-10}~M_{\odot}$yr$^{-1}$.  This mass-loss rate depends on the assumed gas temperature and the wind height. A change of  $T_{\rm gas}$ from 5000~K to 10,000~K would decrease the mass-loss rate by one order of magnitude. 

\subsection{FUV Temporal Variability of CVSO~109 A}
\label{sec:FUV_variability}

The ULLYSES dataset offers a snapshot of the UV radiation field evolution of young stars on timescales of minutes to hours. For CTTS with previous UV observations such as CVSO~109 A, comparing multiple epochs of FUV spectra shows how the UV radiation from TTS changes on multiyear timescales. CVSO~109 A was observed with the COS/G160M mode on 2014 January 1 (program 13363). The H$_{2}$, \ion{C}{4}, and FUV continua were analyzed by~\citet{france17}, who found the \ion{C}{4} luminosity to be among the highest of the young stars in that study. These authors also noted the lack of a ``1600~\AA\ bump" (molecular dissociation continuum) feature. The lack of molecular continuum emission is consistent with an optically thick gas and dust disk that prevents Ly$\alpha$ from dissociating a sufficient quantity of water to produce the molecular features seen in some stars with higher infrared slope ($n_{13-31}$) measurements  ~\citep{france17}.   Figure~\ref{fig:fuv_variability} compares the fluorescent H$_{2}$ line profiles of the (1--7) R(3) Lyman band line at 1489.57~\AA, the \ion{C}{4} doublet, and the FUV continuum emission in the 2014 and 2020 data. The H$_2$ line profiles are similar in flux and FWHM between the two datasets,\footnote{FWHM(H$_2)_{2020}=43.2\pm2.0~\kms$ and F(H$_2)_{2020}=2.60~(\pm0.17)\times10^{-15}~\escm$, compared with FWHM(H$_2)_{2014}=36.3\pm1.1~\kms$ and F(H$_2)_{2014}=2.80~(\pm0.12)\times10^{-15}~\escm$.} suggesting that the Ly$\alpha$ flux and star-disk geometry are similar between the two epochs. 

The underlying FUV continuum level is approximately three times higher in 2020 (1.5 vs.\ $0.5\times10^{-15}~\escm$~\AA\ over 1430--1440~\AA;  Figure~\ref{fig:fuv_variability}, right). Interestingly, while the FUV continuum flux is higher in 2020 compared with 2014, the \ion{C}{4} flux has declined. The 2020-to-2014 flux ratio of the \ion{C}{4} doublet is approximately 0.6: L(\ion{C}{4})$_{2020}=1.06\times10^{30}$ erg~s$^{-1}$ compared with L(\ion{C}{4})$_{2014}=1.84\times10^{30}$ erg~s$^{-1}$. This suggests a surprising anticorrelation between the FUV continuum and \ion{C}{4} fluxes, in contrast with the strong positive correlation seen in flux--flux plots based on a single measurement epoch \citep[Figure~5]{france14}.

\section{Summary and Conclusions} \label{summary}

Here we introduced ODYSSEUS, a multiwavelength survey using the {\it HST} ULLYSES Director’s Discretionary Program of TTS with the goal of studying mass accretion via the stellar magnetic field, mass outflow via winds and jets, and the structure and chemistry of the inner planet-forming disk regions. In addition to the FUV and NUV data provided by ULLYSES, ODYSSEUS will obtain supplemental data from various other instruments, including optical and NIR spectra, optical photometry, and X-ray data for select targets.

Our current work presented initial results of one survey target, CVSO~109, as a demonstration of the science analysis that ODYSSEUS will undertake.

\begin{itemize}

\item We determined the following stellar properties of components CVSO~109A and 109B individually: spectral energy disk, spectral type, equivalent width (H$\alpha$), extinction, stellar luminosity, mass, radial velocity, and inclination.  

\item We presented multiband light curves and H$\alpha$ measurements that show that CVSO~109 displayed significant changes in brightness and accretion during our observations. We concluded that the {\it HST} observations occurred during a local maximum in the light curve corresponding to a modest accretion event.

\item We reproduced the NUV--NIR continuum of CVSO~109A with models of the accretion shock and accretion disk. We measured {\mdot}$\sim3\times10^{-8}$ {\msunyr} and discovered that the inner wall radius is located at $\sim0.1$~au, assuming a dust destruction temperature of 1400~K.

\item The narrow component of the profile of the {\civ} 1549 line in the CVSO~109A spectrum can be reproduced with the accretion shock model from the postshock region. The broad component of the line could arise from hot regions in the magnetospheric flows.

\item Magnetospheric accretion flow modeling of the H$\alpha$ and H$\beta$ emission lines measured {\mdot}$\sim2\times10^{-8}$ {\msunyr} and an inclination angle of $\sim40^{\circ}$.

\item We detected evidence of a fast wind from the inner disk of CVSO~109 in the FUV line profiles of \ion{O}{1}, \ion{Al}{2}, \ion{He}{1}, \ion{Si}{2}, \ion{C}{2}, and \ion{Si}{3}. We estimated a mass-loss rate, $\dot{M}_w$, between $3\times10^{-10}$ to $6\times10^{-9}$\,\msunyr.

\item We measured an FUV luminosity (including the FUV continuum, Ly$\alpha$, and \ion{C}{4}) for CVSO~109 of $6.7\times10^{-13}~\escm$. We found a lower than typical Ly$\alpha$ contribution to the total FUV radiation field, which is unusual given CVSO~109's older age (at which disks are expected to be more settled and therefore have greater Ly$\alpha$ propagation). 

\item We measured an average H$_2$ emitting radius in the range 0.06--0.6~au for CVSO~109, typical of TTS disks and pointing to a gas-rich disk.

\item Compared with archival UV data from 2014, we find that the FUV continuum level is about three times higher in 2020 and the \ion{C}{4} flux has declined by about 40\%, suggesting a surprising anticorrelation between the FUV continuum and \ion{C}{4} fluxes.  

\end{itemize}

CVSO~109 serves as an example of the kind of analysis and results we expect to obtain with ODYSSEUS, which will assemble simultaneous and contemporaneous datasets for each star in the ULLYSES sample. The main advantage over previous works is that ODYSSEUS will consider all of the coordinated observational information together, providing a new, global view of CTTS. For example, self-consistent mass-loss rates and accretion rates will be determined from several probes simultaneously and contemporaneously, which will render the determinations very robust if they coincide. And if inconsistencies arise between the estimates from different datasets, ODYSSEUS will be well suited to identify potential issues and future lines of research. 

An expansive investigation such as ODYSSEUS has never been attempted for a single CTTS and will now become the new standard for the ULLYSES sample. We expect ODYSSEUS will greatly enhance our knowledge of the magnetospheric accretion and ejection processes in CTTS and their high-energy radiation fields. 

\acknowledgments{
This work was supported by {\it HST} AR-16129 and benefited from discussions with the ODYSSEUS team (https://sites.bu.edu/odysseus/). Based on observations obtained with the NASA/ESA Hubble Space Telescope, retrieved from the Mikulski Archive for Space Telescopes (MAST) at the Space Telescope Science Institute (STScI). STScI is operated by the Association of Universities for Research in Astronomy Inc.\ under NASA contract NAS 5-26555. Based on observations collected at the European Southern Observatory under ESO program 106.20Z8. We acknowledge financial support from the project PRIN-INAF 2019 ``Spectroscopically Tracing the Disk Dispersal Evolution (STRADE)." G.J.H.\ is supported by National Key R\&D Program of China No. 2019YFA0405100 and by National Science Foundation of China general grant 11773002.  J.H.\ and J.S.\ acknowledge support from the National Research Council of M\'exico  (CONACyT)  project  No.  86372  and  the  PAPIIT-UNAM project IA102921. FMW acknowledges support through NSF grant 1611443. J.C.W.\ and  A.S.A.\ are supported by the STFC grant number ST/S000399/1. Observing time with SMARTS/CHIRON is made possible by a research support grant from Stony Brook University. This project has received funding from the European Research Council (ERC) under the European Union's Horizon 2020 research and innovation program under grant agreements No.\ 716155 (SACCRED) and No.\ 681601 (BuildingPlanS). This paper utilizes the D’Alessio irradiated accretion disk (DIAD) code. We wish to recognize the work of Paola D’Alessio, who passed away in 2013. Her legacy and pioneering work live on through her substantial contributions to the field. We thank Hodari-Sadiki James, Leonardo Paredes, and Todd Henry for managing the CHIRON spectrograph and for their prompt scheduling of the requested observations. Time on the AAVSOnet is awarded competitively; we acknowledge the contributions of AAVSO to supporting variable star science and thank Ken Menzies for coordinating and scheduling the observations.
We thank the referee for a constructive review.
}

\bibliographystyle{aasjournal}
\bibliography{main.bib}

\allauthors

\end{document}